\documentclass{PoS}

\usepackage{amsmath,amssymb,amsfonts,cite}

\newcommand{\lsim}
{\;\raisebox{-.3em}{$\stackrel{\displaystyle <}{\sim}$}\;}

\newcommand{\gmt}{$(g-2)_\mu$}
\newcommand{\br}{{\rm BR}}
\newcommand{\bsg}{BR($b \to s \gamma$)}
\newcommand{\btn}{BR($B_u \to \tau \nu_\tau$)}
\newcommand{\bmm}{BR($B_s \to \mu^+\mu^-$)}

\newcommand{\Och}{\ensuremath{\Omega_\chi h^2}}

\newcommand{\MW}{M_W}
\newcommand{\MZ}{M_Z}
\newcommand{\Mh}{M_h}
\newcommand{\MA}{M_A}
\newcommand{\MH}{M_H}
\newcommand{\mt}{m_t}
\newcommand{\mgl}{m_{\tilde g}}

\newcommand{\neu}[1]{\tilde \chi^0_{#1}}
\newcommand{\mneu}[1]{m_{\tilde \chi^0_{#1}}}
\newcommand{\mste}{m_{\tilde t_1}}
\newcommand{\mstaue}{m_{\staue}}
\newcommand{\staue}{\tilde \tau_1}
\newcommand{\tb}{\tan\beta}
\newcommand{\ecm}{\sqrt{s}}
\newcommand{\tev}{\,\, \mathrm{TeV}}
\newcommand{\gev}{\,\, \mathrm{GeV}}

\def\reffi#1{\mbox{Fig.~\ref{#1}}}
\def\reffis#1{\mbox{Figs.~\ref{#1}}}
\def\refta#1{\mbox{Tab.~\ref{#1}}}
\def\refse#1{\mbox{Sect.~\ref{#1}}}

\def\citere#1{\mbox{Ref.~\cite{#1}}}
\def\citeres#1{\mbox{Refs.~\cite{#1}}}
\newcommand{\ETslash}{/ \hspace{-.7em} E_T}

\graphicspath{{figs/}}

\title{SUSY Predictions for and from the LHC}

\ShortTitle{SUSY Predictions for and from the LHC}

\author{\speaker{Sven Heinemeyer}\\
        Instituto de F\'isica de Cantabria (CSIC-UC), Santander, Spain\\
        E-mail: \email{Sven.Heinemeyer@cern.ch}}

\abstract{
On the basis of frequentist analyses of experimental constraints
from electroweak precision data, \gmt, $B$ physics and cosmological data, we
predict the masses of Higgs bosons and SUSY particles 
of the CMSSM, NUHM1, VCMSSM and mSUGRA.
In the CMSSM, NUHM1 and VCMSSM we find preferences for
sparticle masses that are relatively light, leading to good prospects
for the LHC and the ILC.
Including first data from SUSY searches at the LHC leads to an
increase of the preferred sparticle mass scales, but improves the
consistency of the model predictions for $\Mh$ with the LEP exclusion
bounds. 
}

\FullConference{Workshop on Discovery Physics at the LHC -Kruger 2010\\
                December 05-10, 2010\\
                Kruger National Park, Mpumalanga, South Africa}

\begin{document}


\section{Introduction}

Supersymmetry (SUSY)~\cite{Nilles:1983ge,Haber:1984rc,Barbieri:1982eh}
is one of the favored ideas for physics beyond
the Standard Model (SM) that is currently explored at the Large Hadron
Collider (LHC). In several recent
papers~\cite{mc1,mc2,mc3,mc35,mc4}, 
we presented results from frequentist analyses of 
the parameter spaces of various GUT based versions of the Minimal
Supersymmetric Standard Model (MSSM), most recently~\cite{mc5} also
including first data from SUSY searches at the LHC~\cite{CMSsusy,ATLASsusy}.
We recall that the CMSSM has four input parameters:  the
universal soft SUSY-breaking scalar and gaugino masses $(m_0, m_{1/2})$,
a universal trilinear soft SUSY-breaking parameter $A_0$ and the ratio of
Higgs v.e.v.'s, $\tb$, as well as the sign of $\mu$ (the magnitude
of $\mu$ and the bilinear SUSY-breaking parameter $B_0$ are fixed by the
electroweak vacuum conditions). The results on
the anomalous magnetic moment of the muon,
\gmt,~\cite{newBNL,g-2,newDavier,newFredl} strongly favor a positive sign.
In the NUHM1, a common soft SUSY-breaking contribution to
the masses of the two Higgs doublets is allowed to vary independently,
so there are five independent parameters. On the 
other hand, the VCMSSM imposes the supplementary
constraint $B_0 = A_0 - m_0$ on the CMSSM (see \citere{mc4} for our
notation), thereby removing $\tb$ as  
a free input and leaving three parameters, on which the further
constraint $m_{3/2} = m_0$ in mSUGRA imposes a severe restriction. 
(Detailed references for the various models can be found in \citere{mc5}.)

Here we review the 
results presented in~\citeres{mc3,mc4,mc5}. 
They include the parameters of the
best-fit points in the CMSSM, NUHM1, VCMSSM and mSUGRA, as well as the 68 and
95\%~CL regions for various sparticle masses applying the
phenomenological, experimental and cosmological constraints. These
include precision electroweak data, \gmt, 
$B$-physics observables (the rates for \bsg\ and \btn,
$B_s$ mixing, and the upper limit on \bmm), the bound on the lightest
MSSM Higgs boson mass, $\Mh$, and the cold dark matter (CDM) density
inferred from astrophysical and cosmological data, assuming that this is
dominated by the relic density of the lightest neutralino, $\Och$.
In the case of \citere{mc5} also recent limits from SUSY searches at
CMS and ATLAS~\cite{CMSsusy,ATLASsusy} are taken into account.


\section{Description of our Frequentist approach}
\label{sec:approach}

We define a global $\chi^2$ likelihood function, which combines all
theoretical predictions with experimental constraints (except the latest
LHC SUSY searches):
\begin{align}
\chi^2 (\equiv \chi^2_{\rm org}) &=
  \sum^N_i \frac{(C_i - P_i)^2}{\sigma(C_i)^2 + \sigma(P_i)^2}
+ \sum^M_i \frac{(f^{\rm obs}_{{\rm SM}_i}
              - f^{\rm fit}_{{\rm SM}_i})^2}{\sigma(f_{{\rm SM}_i})^2}
\nonumber \\[.2em]
&+ {\chi^2(\Mh) + \chi^2(\br(B_s \to \mu\mu))}
\nonumber \\[.2em]
&+ {\chi^2(\mbox{SUSY search limits from LEP and Tevatron})~.}
\label{eqn:chi2}
\end{align} 
Here $N$ is the number of observables studied, $C_i$ represents an
experimentally measured value (constraint) and each $P_i$ defines a
prediction for the corresponding constraint that depends on the
supersymmetric parameters.
The experimental uncertainty, $\sigma(C_i)$, of each measurement is
taken to be both statistically and systematically independent of the
corresponding theoretical uncertainty, $\sigma(P_i)$, in its
prediction. We denote by
$\chi^2(\Mh)$ and $\chi^2(\br(B_s \to \mu\mu))$ the $\chi^2$
contributions from the two measurements for which only one-sided
bounds are available so far, as discussed below.
Furthermore we include the lower limits from the direct searches
for SUSY particles at LEP~\cite{LEPSUSY} as one-sided limits, denoted by 
``$\chi^2(\mbox{SUSY search limits})$'' in eq.~(\ref{eqn:chi2}).

We stress that in \cite{mc3,mc4,mc5} (as in~\cite{mc1,mc2,mc35})
the three SM parameters
$f_{\rm SM} = \{\Delta\alpha_{\rm had}, \mt, \MZ \}$ are included as fit
parameters and allowed to vary with their current experimental
resolutions $\sigma(f_{\rm SM})$. We do not
include $\alpha_s$ as a fit parameter, 
which would have only a minor impact on the analysis.

Formulating the fit in this fashion has the advantage that the
$\chi^2$ probability, $P(\chi^2, N_{\rm dof})$,
properly accounts for the number of degrees of freedom, $N_{\rm dof}$,
in the fit and thus represents a quantitative and meaningful measure for
the ``goodness-of-fit.'' In previous studies \cite{mc1},
$P(\chi^2, N_{\rm dof})$ has been verified to have a flat distribution,
thus yielding a reliable estimate of the confidence level for any particular
point in parameter space. Furthermore, an important aspect of the
formulation is that all model parameters are varied simultaneously in
our MCMC sampling, and care is exercised to fully explore the
multi-dimensional space, including possible interdependencies between
parameters.  All confidence levels for selected model parameters are
performed  by scanning over the desired parameters while  
minimizing the $\chi^2$ function with respect to all other model parameters. 
The function values where $\chi^2(x)$ is found to be equal to 
$\chi^2_{min}+ \Delta \chi^2$ determine the confidence level
contour. For two-dimensional parameter scans we use 
$\Delta \chi^2 =2.23 (5.99)$ to determine the 68\%(95\%) confidence
level contours. 
Only experimental constraints are imposed when deriving confidence level
contours, without any arbitrary or direct constraints placed on model
parameters themselves.
This leads to robust and statistically meaningful
estimates of the total 68\% and 95\% confidence levels,
which may be composed of multiple separated contours.

The experimental constraints used in our analyses are listed in
Table~1 in \cite{mc3}. 
The only significant changes are an updated value of the top quark
mass, $m_t^{\rm exp} = 173.3 \pm 1.1 \gev$~\cite{mt1733} and the 
use of the new $e^+ e^-$ determination of the SM
contribution to \gmt~\cite{newDavier}, 
$a_\mu^{\rm SUSY} = (28.7 \pm 8.0) \times 10^{-10}$,
see also \citere{newFredl}.

One important comment concerns our implementation of the LEP
constraint on $\Mh$. The value quoted in Table~1 of \citere{mc3}, 
$\MH > 114.4 \gev$,
was derived within the SM~\cite{Barate:2003sz}, and is applicable to the
CMSSM, VCMSSM and mSUGRA, in which 
the relevant Higgs couplings are very similar to those in the 
SM~\cite{Ellis:2001qv,Ambrosanio:2001xb}, so that the SM
exclusion results can be used, supplemented with an additional theoretical
uncertainty:
we evaluate the $\chi^2(\Mh)$ contribution within the CMSSM, VCMSSM and
mSUGRA using the
formula
\begin{align}
\chi^2(\Mh) = \frac{(\Mh - \Mh^{\rm limit})^2}{(1.1 \gev)^2 + (1.5 \gev)^2}~,
\label{chi2Mh}
\end{align}
with $\Mh^{\rm limit} = 115.0 \gev$ 
for $\Mh < 115.0 \gev$. 
Larger masses do not receive a $\chi^2(\Mh)$ contribution. 
We use $115.0 \gev$ so as to incorporate
a conservative consideration of experimental systematic effects.
The $1.5 \gev$ in the denominator corresponds to a convolution of
 the likelihood function with a Gaussian function, $\tilde\Phi_{1.5}(x)$,
normalized to unity and centered around $\Mh$, whose width is $1.5 \gev$,
representing the theory uncertainty on $\Mh$~\cite{Degrassi:2002fi}.
In this way, a theoretical uncertainty of up to $3 \gev$ is assigned for 
$\sim 95\%$ of all $\Mh$ values corresponding to one parameter point. 
The $1.1 \gev$ term in the denominator corresponds to a parameterization
of the $CL_s$ curve given in the final SM LEP Higgs
result~\cite{Barate:2003sz}. 

Within the NUHM1 the situation is somewhat more involved, since, for
instance, a strong suppression of the $ZZh$ coupling can occur,
invalidating the SM exclusion bounds. 
In order to find a more reliable 95\% CL exclusion limit for $\Mh$ in the
case that the SM limit cannot be applied, we use the following procedure.
The main exclusion bound from LEP searches comes from the channel
$e^+e^- \to ZH, H \to b \bar b$. The Higgs boson mass limit in this
channel is given as a function of the $ZZH$ coupling in~\cite{Schael:2006cr}. 
A reduction in the $ZZh$ coupling in the
NUHM1 relative to its SM value can be translated into a lower
limit on the lightest NUHM1 Higgs mass, $\Mh^{{\rm limit},0}$, shifted
to lower values with respect to the SM limit of $114.4 \gev$. (The
actual number is obtained using the code {\tt HiggsBounds}~\cite{higgsbounds}
that incorporates the LEP (and Tevatron) limits on neutral Higgs boson
searches.) 
For values of $\Mh \lsim 86 \gev$ the reduction of the $ZZh$
couplings required to evade the LEP bounds becomes very strong, and we
add a brick-wall contribution to the $\chi^2$ function below this value
(which has no influence on our results).
Finally, eq.~(\ref{chi2Mh}) is used with 
$\Mh^{\rm limit} = \Mh^{{\rm limit},0} + 0.6 \gev$ to ensure a smooth
transition to the SM case, see~\cite{mc3} for more details.

To include the recent LHC searches for SUSY we calculate~\cite{mc5}
\begin{align}
\chi^2_{\rm ATLAS} &= \chi^2_{\rm org} 
 + \Delta\chi^2_{\rm ATLAS}~, \nonumber \\
\chi^2_{\rm CMS} &= \chi^2_{\rm org} 
 + \Delta\chi^2_{\rm CMS}~,
\label{eq:chi2new}
\end{align}
i.e.\ we investigate the impact of the SUSY searches at ATLAS and CMS
{\em separately} and do not attempt a combination of these searches.

The CMS result~\cite{CMSsusy} is based on a search for 
multijet + $\ETslash$ events without accompanying leptons. 
The 13 events found in the signal region were compatible with the $\sim 10.5$
expected from SM backgrounds with a probability value of 30\% 
The observed result allowed CMS to set a 95\%~CL (i.e.,
$1.96\,\sigma$) upper limit of 13.4 signal events. This would correspond 
to $2.5 \pm (13.4 - 2.5)/1.96 = 2.5 \pm 5.6$ events for any possible
signal, yielding $\chi^2_{\infty,{\rm CMS}} = 0.85$ for large sparticle masses.
The central CMS result, shown in Fig.~5 of ~\cite{CMSsusy}, is a 95\%~CL
exclusion contour in the $(m_0, m_{1/2})$ plane of the CMSSM for the
particular values $\tb = 3, A_0 = 0$ and $\mu > 0$. However, the
sensitivity of a search for multijet + $\ETslash$ events
is largely independent of these additional parameters within the
CMSSM~\cite{CMSsusy}, and can also be taken over to the NUHM1, VCMSSM and
mSUGRA models, which have similar signatures in these search channels.

Fig.~5 of~\cite{CMSsusy} also presents a $(m_0, m_{1/2})$ contour
for the 95\% CL exclusion expected in the absence of any signal, corresponding
to $5.56 \times 1.96 = 10.9$ events.  This contour 
would correspond to an {\it apparent} significance
of $(10.9 - 2.5)/5.56 \sim 1.5\,\sigma$ and hence $\Delta \chi^2 \sim 4$.
The observed 95\% CL contour, on the other hand,
corresponds to $\Delta \chi^2 = 5.99$.
We approximate the impact of the new CMS constraint by 
$\Delta \chi^2_{\rm CMS} \sim \chi^2_{\infty,{\rm CMS}} |(M_C/M) - 1|^{-p_C}$ 
(where $M \equiv \sqrt{m_0^2 + m_{1/2}^2}$) for each ray in the $(m_0, m_{1/2})$ 
plane, fitting the parameters $M_C, p_C$ by requiring 
$\Delta \chi^2 \sim 4, 5.99$ on the expected and observed 95\%
exclusion contours shown in Fig.~5 of~\cite{CMSsusy}, see \citere{mc5}
for details.

The ATLAS result~\cite{ATLASsusy} is based on a search for 
multijet + $\ETslash$ events with one accompanying electron or muon. The 2
events found in the signal region were compatible with the $\sim 4.1$
expected from SM backgrounds with a probability value of 16\%.
The central ATLAS result, shown in Fig.~2 of ~\cite{ATLASsusy}, 
is again a 95\%~CL exclusion contour in
the $(m_0, m_{1/2})$ plane of the CMSSM for the
particular values $\tb = 3, A_0 = 0$ and $\mu > 0$, which is also
only moderately dependent on these additional parameters within the
CMSSM~\cite{ATLASsusy}, and can also be taken over to the NUHM1, VCMSSM and
mSUGRA models. We construct $\Delta \chi^2_{\rm ATLAS}$ in a similar
fashion as for CMS, where all the details can be found in \citere{mc5}.

The numerical evaluation of the frequentist likelihood function
using the constraints has been performed with the 
{\tt MasterCode}~\cite{mc1,mc2,mc3,mc35,mc4,mc5,mc-web},
which includes the following theoretical codes. For the RGE running of
the soft SUSY-breaking parameters, it uses
{\tt SoftSUSY}~\cite{Allanach:2001kg}, which is combined consistently
with the codes used for the various low-energy observables. 
At the electroweak scale we have included various codes:
{\tt FeynHiggs}~\cite{Degrassi:2002fi,Heinemeyer:1998np,Heinemeyer:1998yj,Frank:2006yh}  
is used for the evaluation of the Higgs masses and
$a_\mu^{\rm SUSY}$ (see also
\cite{Moroi:1995yh,Degrassi:1998es,Heinemeyer:2003dq,Heinemeyer:2004yq}).
For flavor-related observables we use 
{\tt SuFla}~\cite{Isidori:2006pk,Isidori:2007jw} as well as 
{\tt SuperIso}~\cite{Mahmoudi:2008tp,Eriksson:2008cx}, and
for the electroweak precision data we have included 
a code based on~\cite{Heinemeyer:2006px,Heinemeyer:2007bw}.
Finally, for dark-matter-related observables, 
{\tt MicrOMEGAs}~\cite{Belanger:2006is,Belanger:2001fz,Belanger:2004yn} and
{\tt DarkSUSY}~\cite{Gondolo:2005we,Gondolo:2004sc} 
have been used.
We made extensive use of the SUSY Les Houches
Accord~\cite{Skands:2003cj,Allanach:2008qq} 
in the combination of the various codes within the {\tt MasterCode}.


\section{Results for Sparticle Masses (without LHC data)}

\begin{figure*}[htb!]
\begin{center}
\resizebox{11.5cm}{!}{\includegraphics{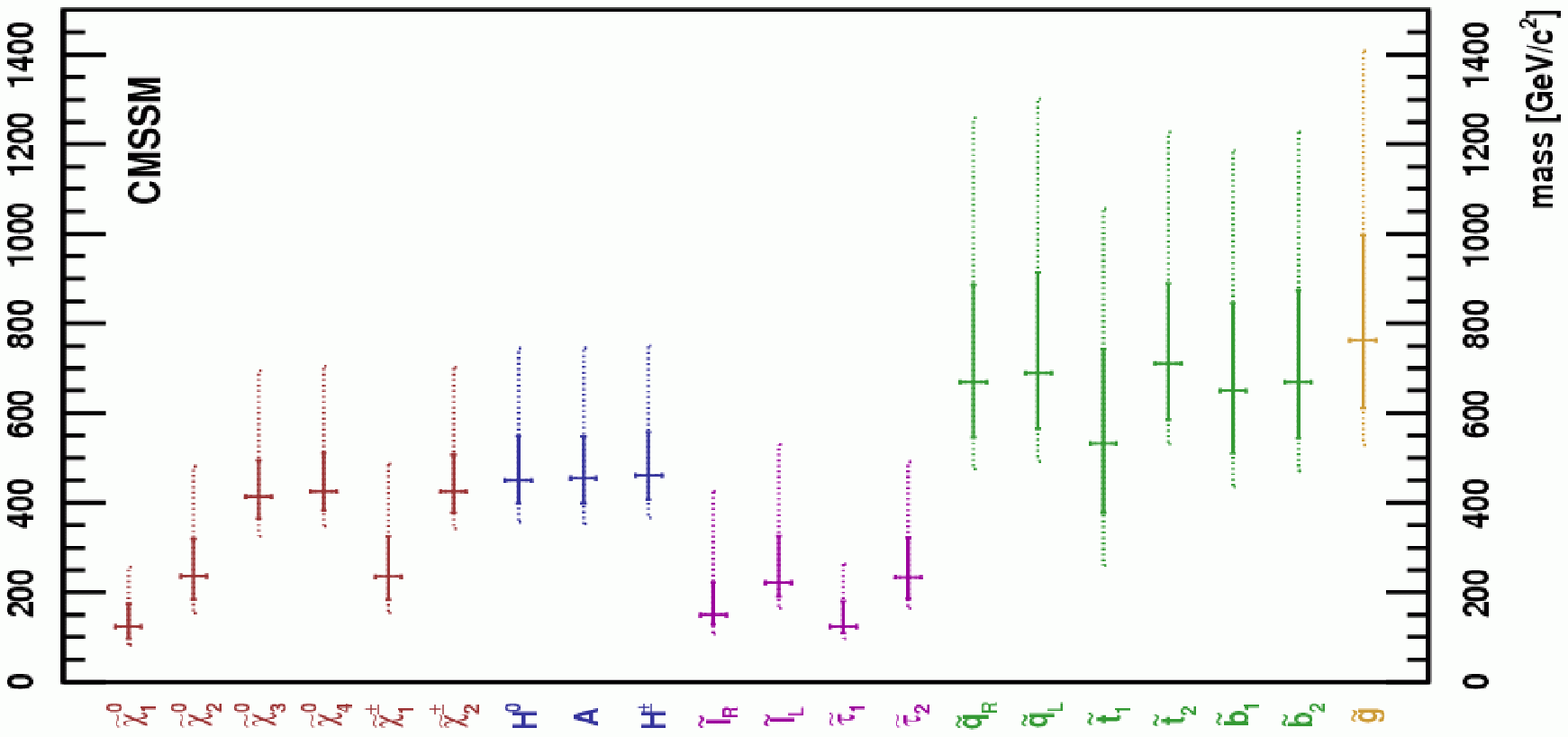}}
\resizebox{11.5cm}{!}{\includegraphics{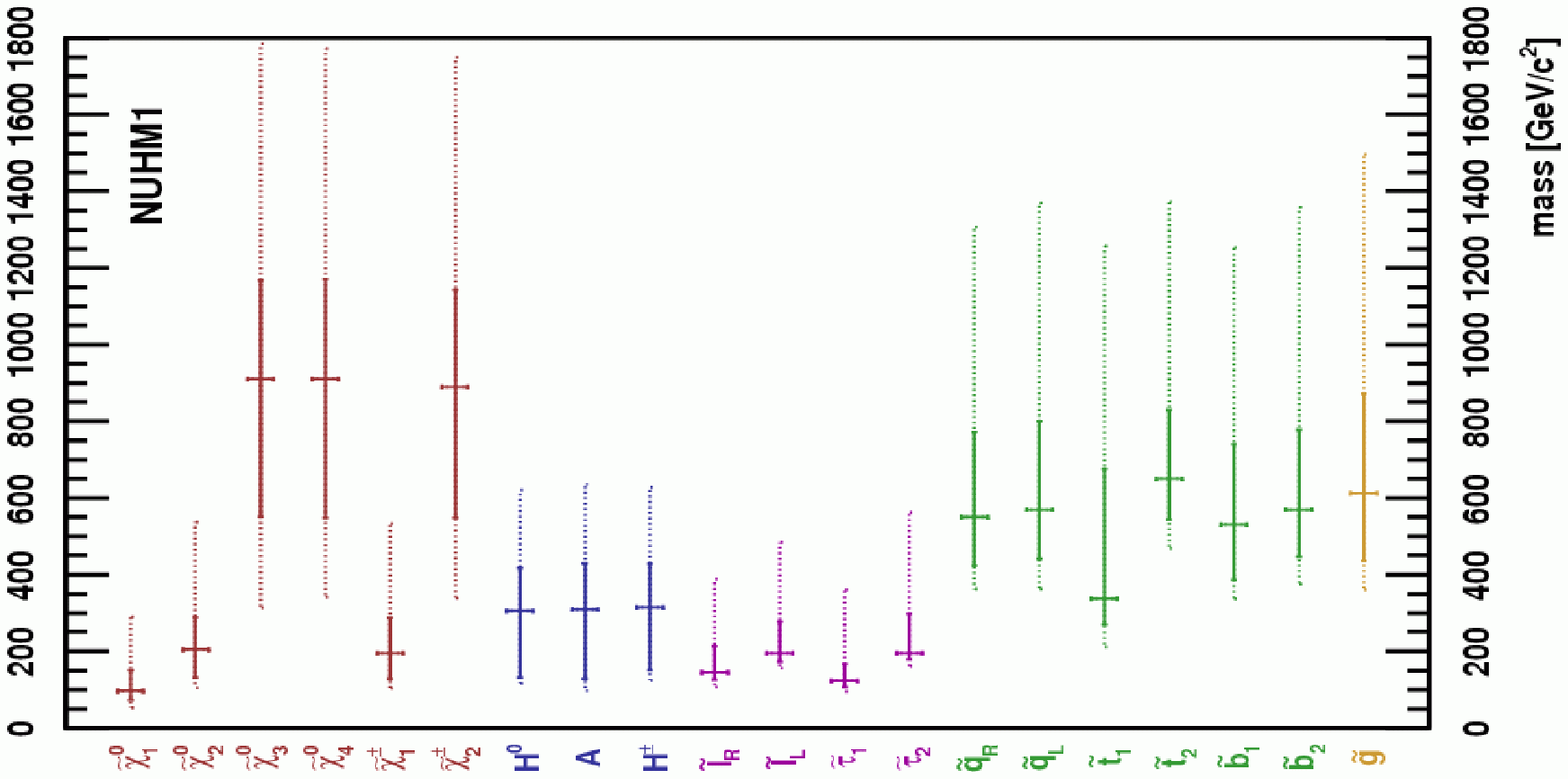}}
\end{center}
\caption{\it Spectra in the CMSSM (upper plot) and the NUHM1
(lower plot)~\cite{mc3}. The horizontal
solid lines indicate the best-fit values, the vertical solid lines
are the 68\% CL
ranges, and the vertical dashed lines are the 95\% CL ranges for the
indicated mass parameters. 
}
\label{fig:spectrum}
\end{figure*}

\begin{figure*}[htb!]
\begin{center}
\resizebox{10.0cm}{!}{\includegraphics{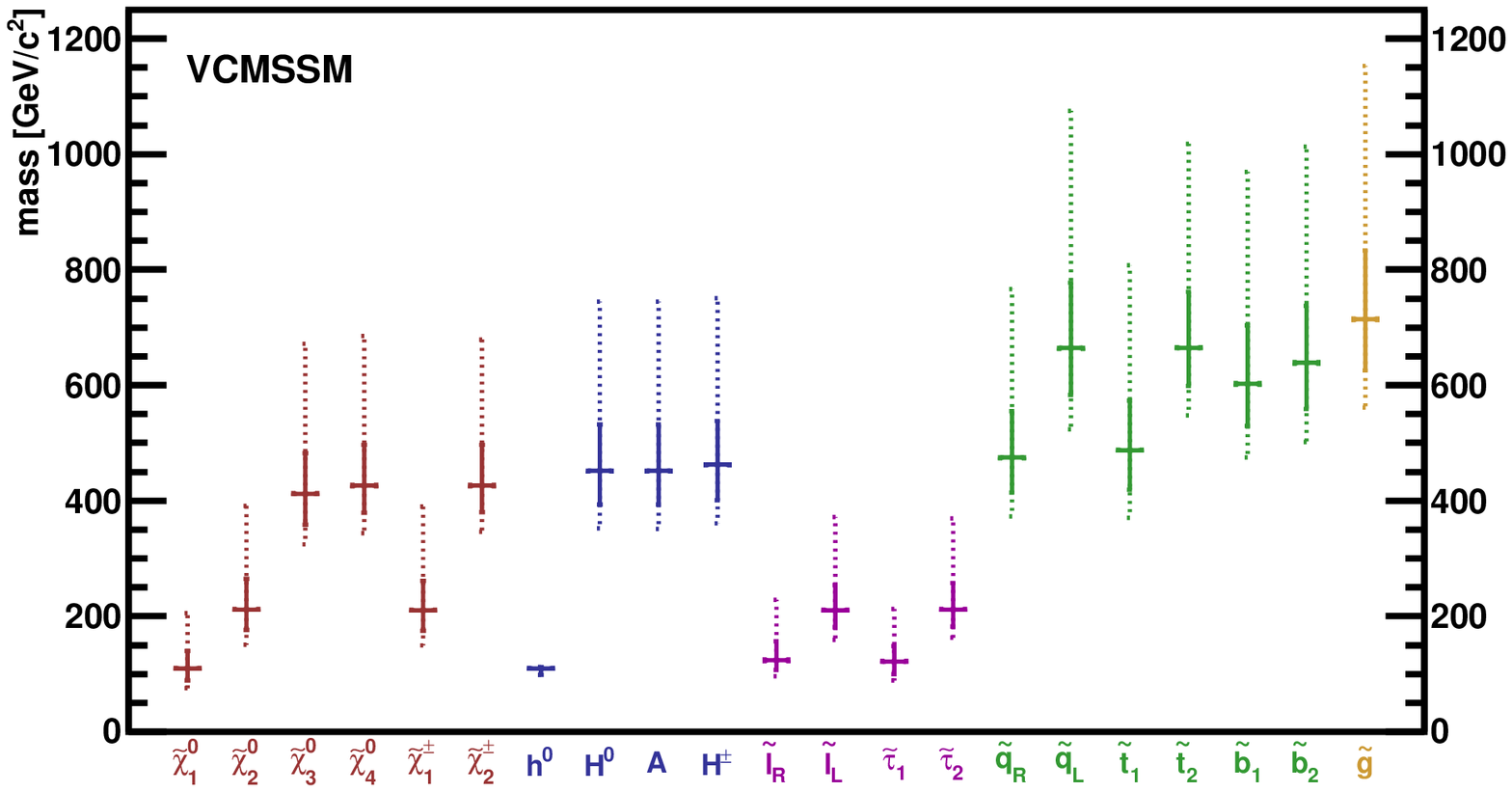}} \\
\resizebox{10.0cm}{!}{\includegraphics{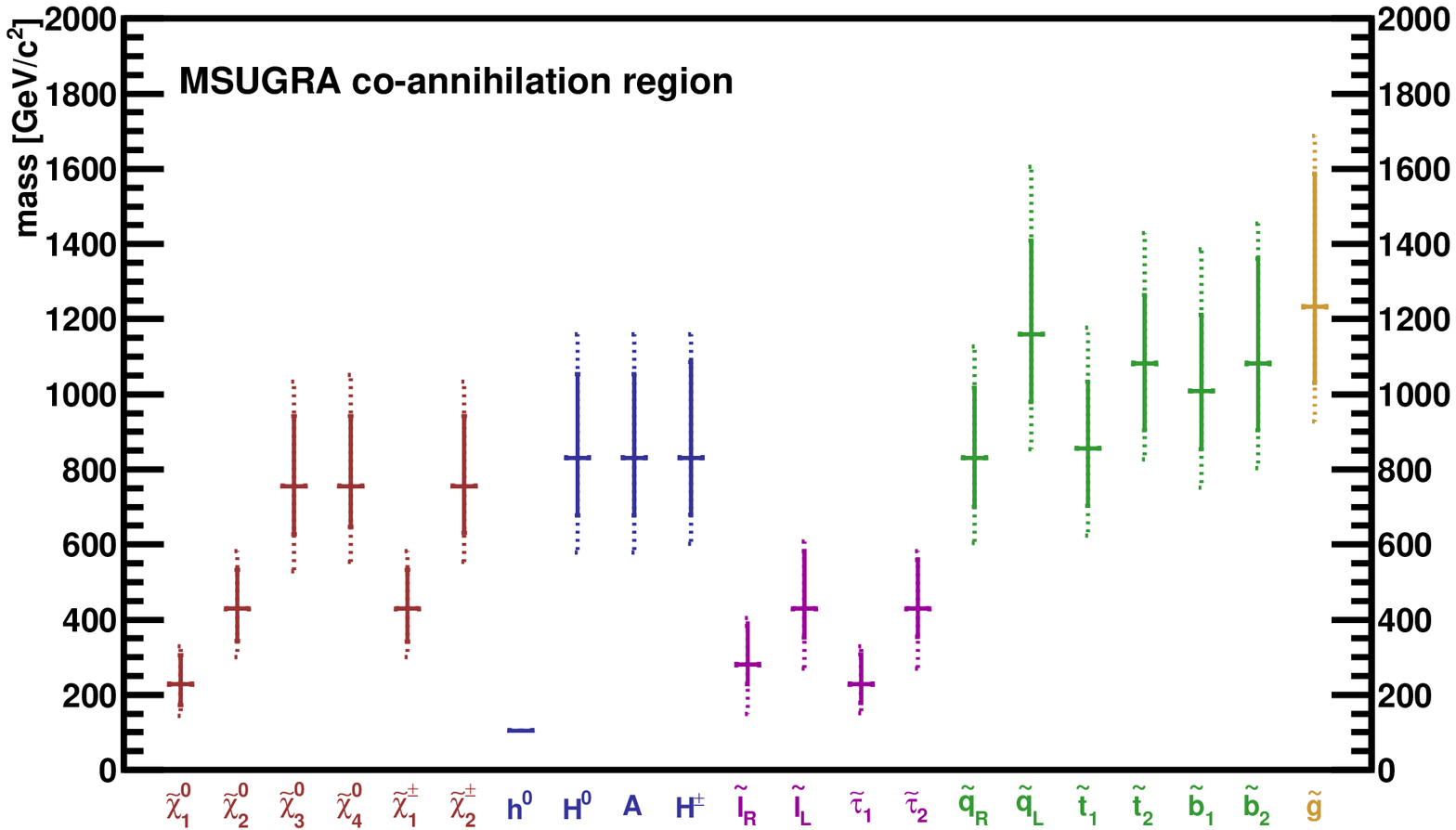}} \\
\resizebox{10.0cm}{!}{\includegraphics{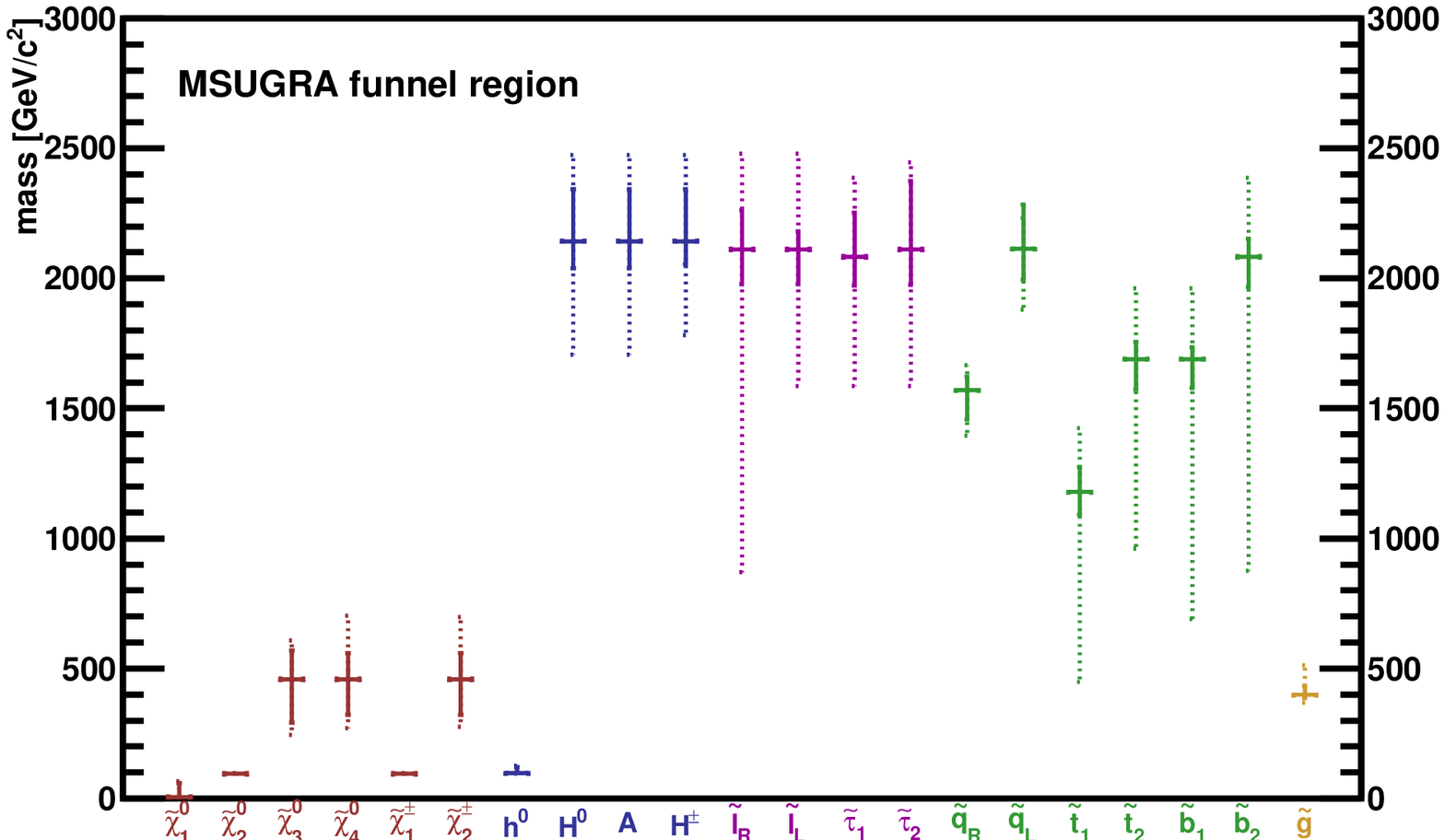}} \\
\end{center}
\caption{\it Spectra in the VCMSSM (top), and mSUGRA in the
coannihilation region (middle) and the funnel region (bottom),
  implementing all the constraints including that on \Och. The horizontal
solid lines indicate the best-fit values, the vertical solid lines
are the 68\% CL ranges, and the vertical dashed lines are the 95\% CL
ranges for the indicated mass parameters. 
}
\label{fig:spectrum2}
\end{figure*}

The best-fit points and respective probabilities are summarized in
\refta{tab:compare} in \refse{sec:lhc}.
Here we review the results for the predictions of sparticles masses in the
CMSSM, NUHM1, VCMSSM and mSUGRA. 
The results for the CMSSM 
spectrum are shown in the upper plot, and for the NUHM1 in the lower
plot of \reffi{fig:spectrum}, whereas the spectra in the VCMSSM can be
found in top plot and mSUGRA in the middle and bottom plot of
\reffi{fig:spectrum2}. The first (middle) mSUGRA plot corresponds to the
overall minimum in $\chi^2 \sim 29$ in this model. 
However, a second local minimum of $\chi^2 \sim 33$
(and hence a ``small'' area allowed at the 95\% CL) can be found along
the light Higgs rapid-annihilation strip with small $m_{1/2}$ are very
large $m_0$. 
The masses shown in the second (lower) plot
correspond to this secondary minimum.

We start our discussion with the gluino mass, $\mgl$.
In both the CMSSM and the NUHM1, the
best-fit points have relatively low values of $\mgl \sim 750$ and $\sim
600 \gev$, respectively. These favored values are well within the range
even of the early operations of the LHC with reduced center-of-mass
energy and limited luminosity. However, even quite large values of 
$\mgl \lsim 2.5 \tev$
are allowed at the 3\,$\sigma$ ($\Delta \chi^2 = 9$) level 
(not shown in Fig.~\ref{fig:spectrum}). The LHC
should be  able to discover a gluino with $\mgl \sim 2.5 \tev$ with
100/fb of integrated luminosity at 
$\ecm = 14 \tev$~\cite{atlastdr,cmstdr}, and the proposed 
sLHC luminosity upgrade to 1000/fb of integrated luminosity at 
$\ecm = 14 \tev$ should permit the discovery of a gluino with 
$\mgl \sim 3 \tev$~ \cite{Gianotti:2002xx}. 
However, Fig.~\ref{fig:spectrum} does demonstrate that, whilst there are good
prospects for discovering SUSY in early LHC running~\cite{mc2}, 
this cannot be `guaranteed'.

The central values of the masses of the supersymmetric
partners of the $u, d, s, c, b$ quarks are slightly lighter than the
gluino, as seen in 
Fig.~\ref{fig:spectrum}. The difference between the gluino and the
squark masses is sensitive primarily to $m_0$.
The reason is that the preferred regions of the
parameter space in both the CMSSM
and the NUHM1 are in the $\neu{1}$-slepton coannihilation
region~\cite{mc2,mc3} where $m_0 < m_{1/2}$. Here $m_0$ makes only
small contributions to the central values of the
squark masses. The SUSY partners of the left-handed components of
the four lightest quarks, the ${\tilde q_L}$, 
are predicted to be slightly heavier than the corresponding right-handed
squarks, ${\tilde q_R}$, as seen by comparing the mass ranges in
Fig.~\ref{fig:spectrum}. As in the case of the gluino,
squark masses up to $\sim 2.5 \tev$ are allowed at the 3\,$\sigma$
level. Comparing 
the upper and lower panels, we see that the squarks are predicted to be
somewhat lighter in the NUHM1 than in the CMSSM, 
but this difference is small compared
with the widths of the corresponding likelihood functions.

Turning now to the likelihood functions for the mass of the lighter stop, 
$\mste$, we find that it is shifted to values somewhat lower than for
the other squark flavors. It can also be seen
that the 2\,$\sigma$ range of its likelihood function differ from those
of the gluino and the other squarks, reflecting the importance of scalar
top mixing. We recall that this depends strongly on the trilinear
soft SUSY-breaking parameter $A_t$ and the
Higgs mixing parameter $\mu$, as well as on the precise value
of $\mt$. 

In the case of the lighter stau $\staue$, see its range in
Fig.~\ref{fig:spectrum}, the mass is very similar to
that of the LSP $\neu{1}$ in the coannihilation region, but this is not
the case in the rapid-annihilation $H, A$ funnel region,
see~\cite{mc3} for details.
In the case of the NUHM1 rapid annihilation is possible also for low
$\tb$, leading to larger values of $m_0$ than in the CMSSM also for
relatively small values of $\mstaue$.

The scalar taus as well as the other scalar leptons are expected to be
relatively light, as can be seen in Fig.~\ref{fig:spectrum}. They would
partially be in the reach of the ILC(500) (i.e.\ with $\sqrt{s} = 500 \gev$)
and at the 95\% CL nearly all be in the reach of the
ILC(1000)~\cite{teslatdr,Brau:2007zza}. 
This also holds for the two lighter neutralinos and the light chargino.

In the case of the VCMSSM, the spectrum is
qualitatively similar to those in the CMSSM and NUHM1.
The two mSUGRA spectra are significantly different from each other and from the
VCMSSM, CMSSM and NUHM1. This is because the coannihilation region has 
$m_{1/2}$ significantly larger than in the
other models, whereas the funnel region has a significantly smaller and very
well-defined value of $m_{1/2}$ and relatively large values of
$m_0$. This bimodality affects directly the preferred values of
$\mneu{1}$ and $\mgl$, and affects the 
other sparticle masses via renormalization effects.
These spectra show that the colored particles are well within
the reach of the LHC for the VCMSSM and mSUGRA  in the coannihilation
region, whereas more integrated luminosity would be
necessary for mSUGRA in the funnel region (except for gluino production). 
In each scenario some SUSY
particles should be accessible at an $e^+e^-$ collider, even with a
center-of-mass energy as low as $500 \gev$.


\section{Prediction for the Higgs sector (without LHC data)}

\begin{figure*}[htb!]
\resizebox{8cm}{!}{\includegraphics{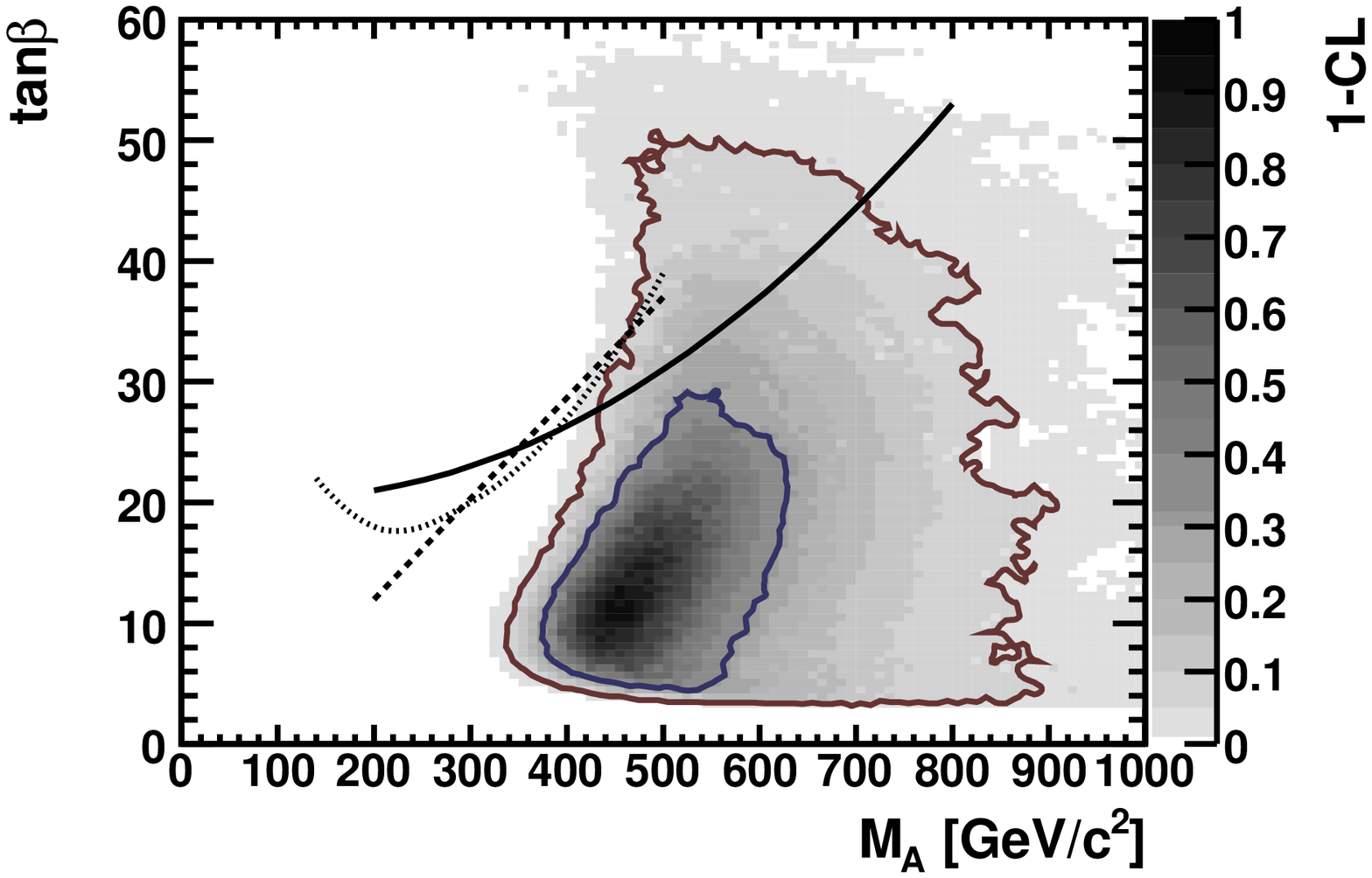}}
\resizebox{8cm}{!}{\includegraphics{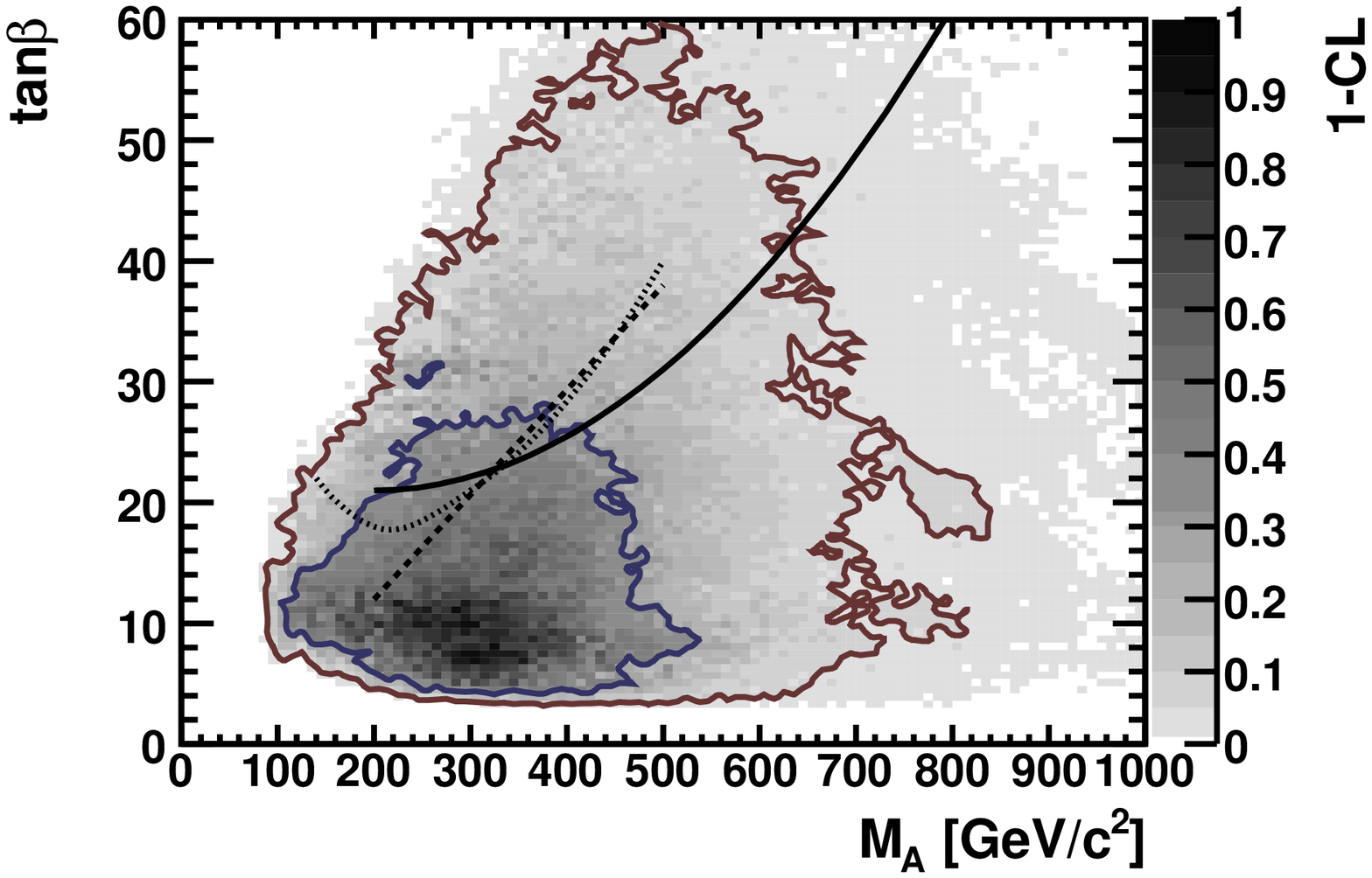}}
\vspace{-2em}
\caption{\it The correlations between $\MA$ and $\tb$
in the CMSSM (left panel) and in the NUHM1 (right panel)~\cite{mc3}.
Also shown are the 5\,$\sigma$ discovery contours 
for observing the heavy MSSM Higgs bosons $H, A$ 
in the three decay channels
$H,A \to \tau^+\tau^- \to {\rm jets}$ (solid line), 
$\rm{jet}+\mu$ (dashed line), $\rm{jet}+e$ (dotted line)
at the LHC. The discovery contours have been obtained using an 
analysis that assumed 30 or 60~fb$^{-1}$ 
collected with the CMS detector~\cite{cmstdr,cmsHiggs}.
}
\label{fig:mAtb}
\vspace{3em}
\end{figure*}

In Fig.~\ref{fig:mAtb} we display the favored regions in the
$(\MA, \tb)$ planes 
for the CMSSM and NUHM1 (taken from \citere{mc3}). 
(Predictions for $\Mh$ in the four models are reviewed in \refse{sec:MhLHC}.)
We see that they are broadly
similar, with little correlation between the two parameters.
Concerning $\tb$, one can observe that while the best fit values lie at
$\tb \approx 11$, the 68 (95)\% CL areas reach up to 
$\tb \approx 30 (50$-$60)$. 
The existing heavy neutral Higgs discovery analyses (performed in the various
benchmark scenarios~\cite{Ellis:2007aa,Ellis:2007ka,benchmark2,benchmark3}) 
cannot directly be applied to the $(\MA, \tb)$ planes in
Fig.~\ref{fig:mAtb}. In order to assess the prospects for discovering
heavy Higgs bosons at the LHC in this context,
we follow the analysis in~\cite{cmsHiggs}, which assumed 30
or 60~fb$^{-1}$ collected with the CMS detector. For evaluating the
Higgs-sector observables including higher-order corrections we use 
the soft
SUSY-breaking parameters of the best-fit points in the CMSSM and the
NUHM1, respectively. We show in Fig.~\ref{fig:mAtb} the 5\,$\sigma$
discovery contours for the three decay channels 
$H,A \to \tau^+\tau^- \to {\rm jets}$ (solid lines), $\rm{jet}+\mu$ (dashed
lines) and $\rm{jet}+e$ (dotted lines).
The parameter regions above and to the left of the curves are within reach 
of the LHC with about 30~fb$^{-1}$ of integrated luminosity.
We see that most of the highest-CL regions lie beyond this reach,
particularly in the CMSSM.
At the ILC(1000) masses up to $\MA \lsim 500 \gev$ can be probed. 
Within the CMSSM this includes the best-fit point, and within the NUHM1
nearly the whole 68\% CL area can be covered.

\section{Results including LHC data}
\label{sec:lhc}

In this section we review the results obtained {\em including} the
recently published SUSY search limits from CMS~\cite{CMSsusy} and 
ATLAS~\cite{ATLASsusy} (taken from \citere{mc5}).%
\footnote{
Anticipated results including early LHC data in the context of
frequentist fits were shown, e.g., in \citeres{mcproc1,mcproc2}.
}%
~In \refta{tab:compare} we compare the best-fit points
found incorporating the CMS or ATLAS constraints~\cite{mc5} with pre-LHC
results~\cite{mc2,mc3,mc4} in the four scenarios. In addition to the
minimum value of $\chi^2$ and the fit probability in each scenario, we include
the values of $m_{1/2}, m_0$, $A_0$ and $\tb$ at  all the best-fit
points, as well as $\Mh$ (discarding the LEP limits).%
\footnote{In \citere{mc5} a systematic error of $\sim 10 (20)\%$ was
estimated in the values for $m_{1/2}$ quoted in Table~\ref{tab:compare} 
at the best-fit points, associated
with ambiguity in the implementation of the CMS (ATLAS) constraint and
the slow variations in the $\chi^2$ functions.}

\begin{table*}[!tbh]
\renewcommand{\arraystretch}{1.2}
\begin{center}
\begin{tabular}{|c||c|c|c|c|c|c||c|c|} \hline
Model & Minimum $\chi^2$ & Probability & $m_{1/2}$ & $m_0$ & $A_0$ & $\tb$ & $\Mh$ (no LEP) \\
& & & (GeV) & (GeV) & (GeV) & & (GeV)\\ \hline \hline
CMSSM      & (21.3) & (32\%) & (320) & (60) & (-170) & (11) & (107.9) \\
with CMS   &  22.0  &   29\%    &  370  &  80  &   -340  &  14  &   112.6 \\
with ATLAS &  24.9  &   16\%    &  400  & 100  &   -430  &  16  &   112.8 \\
\hline
NUHM1      & (19.3) & (31\%) & (260) & (110) & (1010) & (8) & (121.9) \\
with CMS   &  20.9  &   28\%    &  380  &   90  &   70  & 14  &  113.5 \\
with ATLAS &  23.3  &  18\%    &  490  &  110  &   -630  & 25  &  116.5 \\
\hline
VCMSSM     & (22.5) & (31\%) & (300) &  (60) &   (30) & (9) & (109.3) \\
with CMS   &  23.8  &   25\%    &  340  &   70  &   50  &  9  &  115.5 \\
with ATLAS &  27.1  &   13\%    &  390  &   90  &   70  & 11  &  117.0 \\
\hline
mSUGRA     & (29.4) & (6.1\%) & (550) & (230)  & (430) & (28) & (107.8) \\
with CMS   &   29.4  &     6.1\%    & 550  & 230 & 430 & 28 &  121.2 \\
with ATLAS &  30.9  &   5.7\%      & 550  & 230 & 430 & 28 &  121.2 \\
\hline
\end{tabular}
\caption{\it Comparison of the best-fit points found previously in the
CMSSM, the NUHM1, the VCMSSM and the coannihilation region of mSUGRA
when the LHC constraints were not included (in
parentheses)~\cite{mc2,mc3,mc4}, and the results of this 
paper incorporating the CMS~\cite{CMSsusy} and ATLAS~\cite{ATLASsusy}
constraints~\cite{mc5}. In addition to the minimum
value of $\chi^2$ and the fit probability in each scenario, we include the
values of $m_{1/2}, m_0, A_0$ and $\tb$ at  all the best-fit points, as well
as the predictions for $\Mh$ {\it neglecting} the LEP constraint.} 
\label{tab:compare}
\end{center}
\end{table*}

\begin{figure*}[htb!]
\resizebox{8cm}{!}{\includegraphics{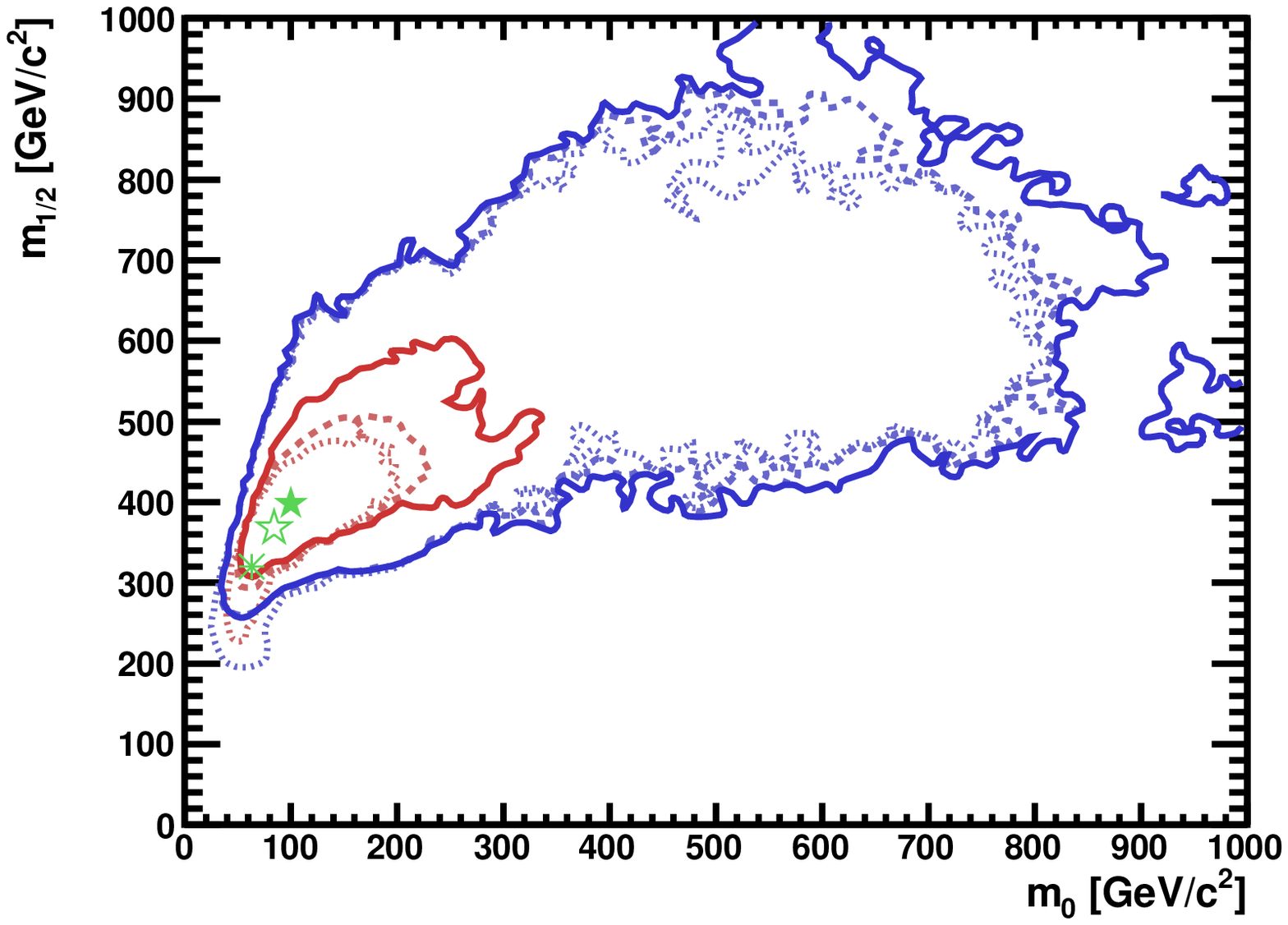}}
\resizebox{8cm}{!}{\includegraphics{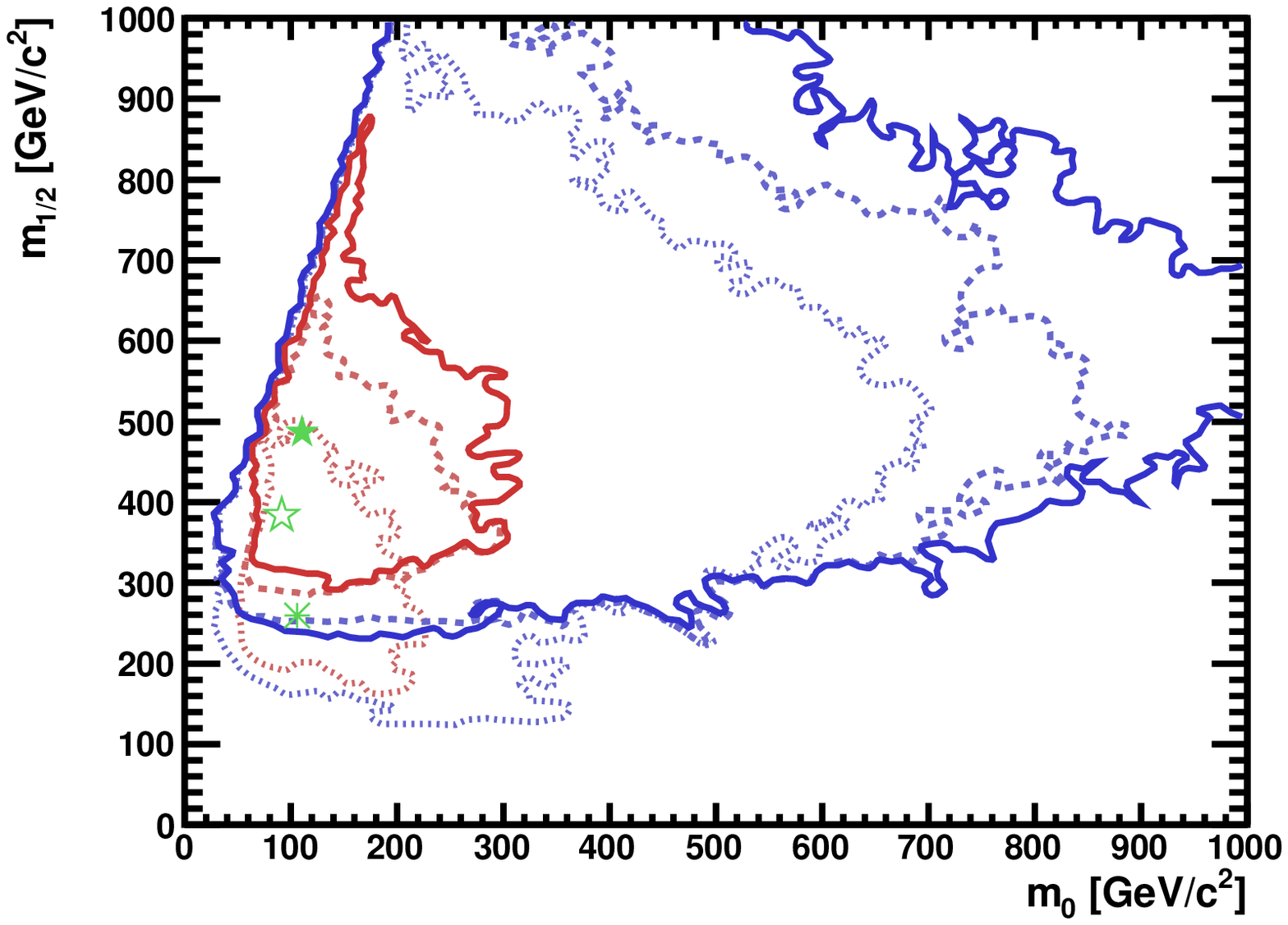}}
\resizebox{8cm}{!}{\includegraphics{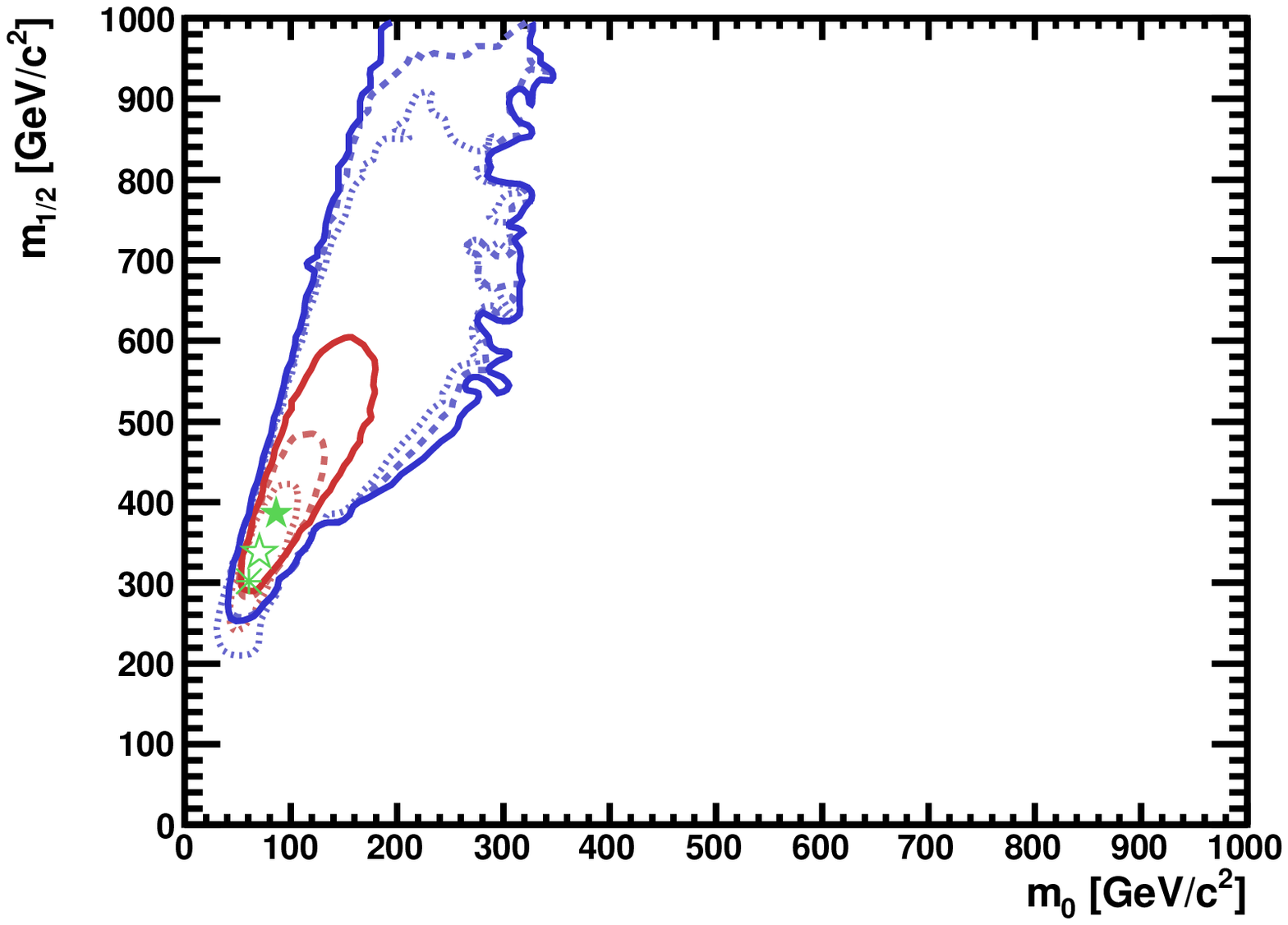}}
\resizebox{8cm}{!}{\includegraphics{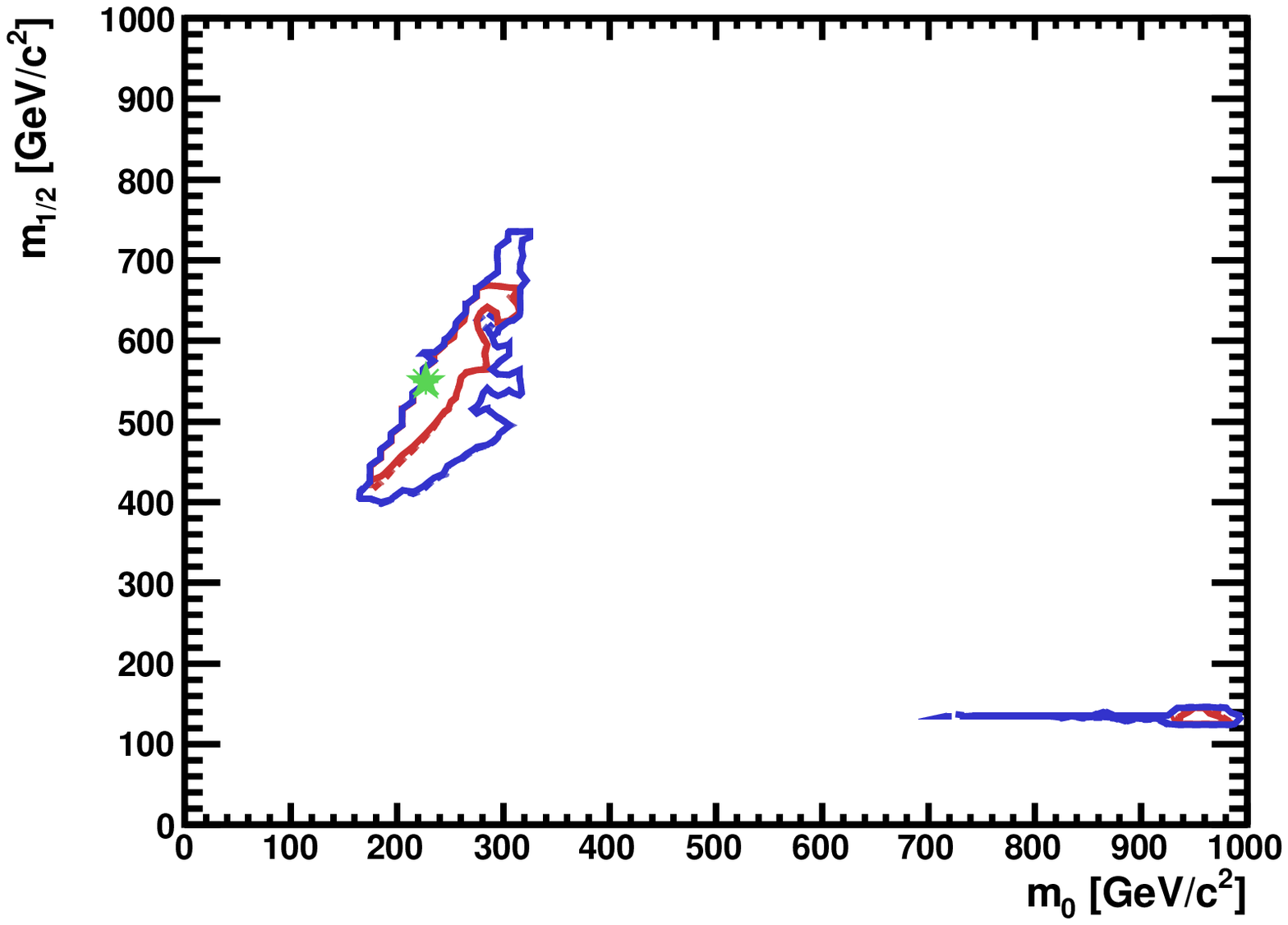}}\\
\vspace{-1cm}
\caption{\it The $(m_0, m_{1/2})$ planes in the CMSSM (upper left),
NUHM1 (top right), VCMSSM (lower left) and mSUGRA (lower right).
In each panel, we show  the 68 and 95\%~CL contours (red and blue,
respectively) both after applying the
CMS~\cite{CMSsusy} and ATLAS~\cite{ATLASsusy}
constraints (dashed and solid lines, 
respectively) and beforehand (dotted lines). Also shown as open
(solid) green stars are the best-fit points found after applying the 
CMS (ATLAS) constraints in each model (see text), and as green
`snowflakes' the previous best-fit points.
}
\label{fig:m0m12}
\end{figure*}

The absence of a supersymmetric signal in the LHC
data~\cite{CMSsusy,ATLASsusy} 
invalidates portions of the CMSSM, NUHM1 and VCMSSM parameter spaces
at low $m_{1/2}$ that were previously favored at the 95\% and 68\%~CL,
but does not impinge significantly on the corresponding regions for
mSUGRA. In the cases 
of the CMSSM and VCMSSM, the LHC data disfavor the low-$m_{1/2}$ tips of
the coannihilation regions and increase visibly the best-fit values of
$m_{1/2}$, as seen in \reffi{fig:m0m12} (and in \refta{tab:compare}).
However, it should be kept in mind that the area around the best-fit
points is very shallow in $\chi^2$, particularly as a function of $m_{1/2}$.
In the case of the NUHM1, the CMS or ATLAS data disfavor a slice of
parameter space at low $m_{1/2}$ and $m_0 < 400 \gev$ extending between the
coannihilation region and the light-Higgs funnel discussed in~\cite{mc4}.
However, as noted in Table~\ref{tab:compare}, the new LHC constraints
do not increase significantly the value of the global $\chi^2$
function for the previous best-fit 
points in any of the models. Applying the CMS constraint we find an
increase of $\Delta \chi^2 \sim 1$. The ATLAS constraint leads to
somewhat stronger increase up to $\Delta\chi^2 \sim 4.5$.
This indicates that there is no significant tension between the CMS
and previous constraints in the contexts of the models studied here, 
and only a mild tension with the new ATLAS limits.
Absence of a SUSY signal at the LHC with a luminosity of $\ge 1$/fb
at a center-of-mass energy $\ge 7$ TeV would be required to increase
the global minimum of $\chi^2$ sufficiently to put pressure on
these models.

\begin{figure*}[htb!]
\resizebox{8cm}{!}{\includegraphics{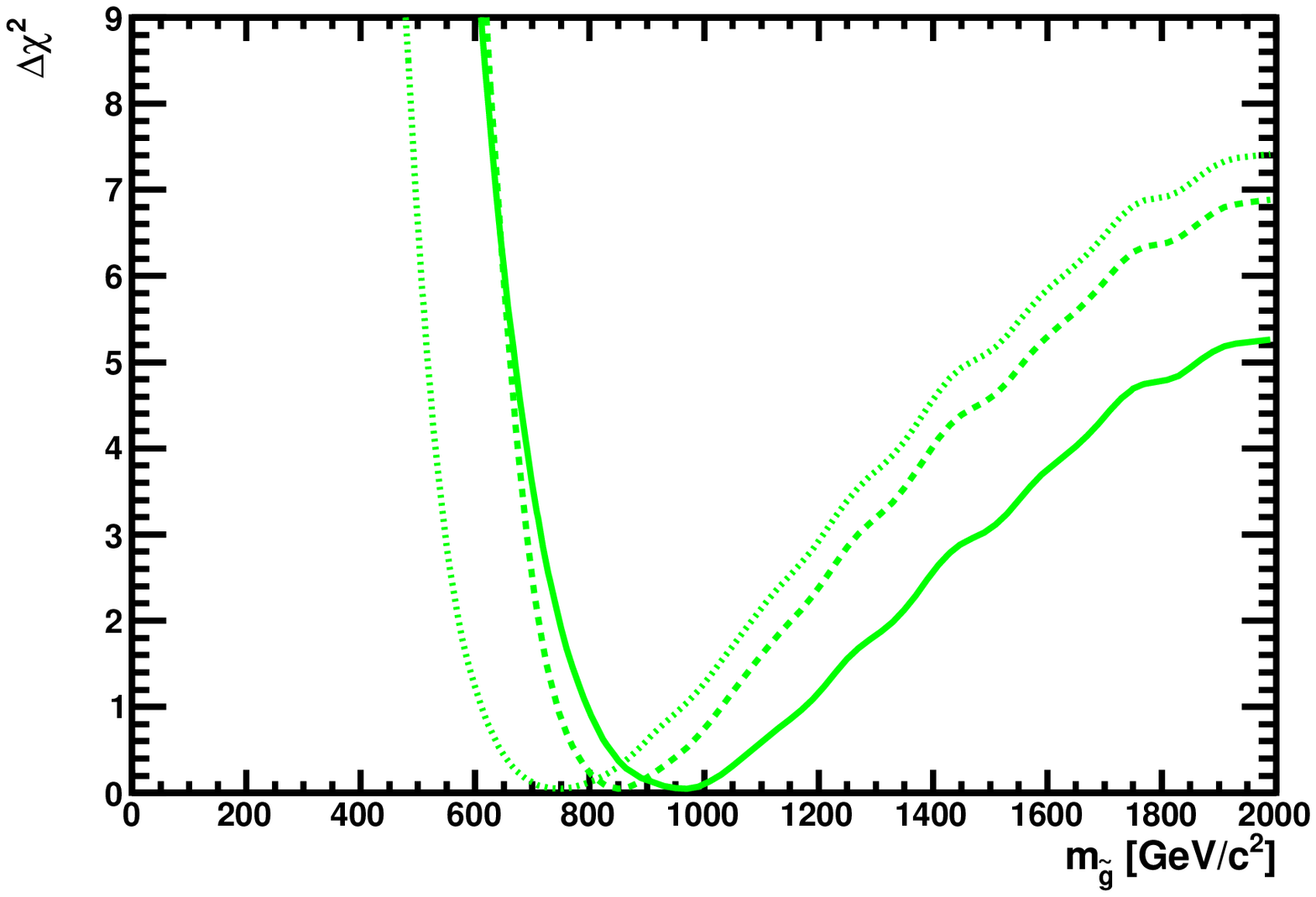}}
\resizebox{8cm}{!}{\includegraphics{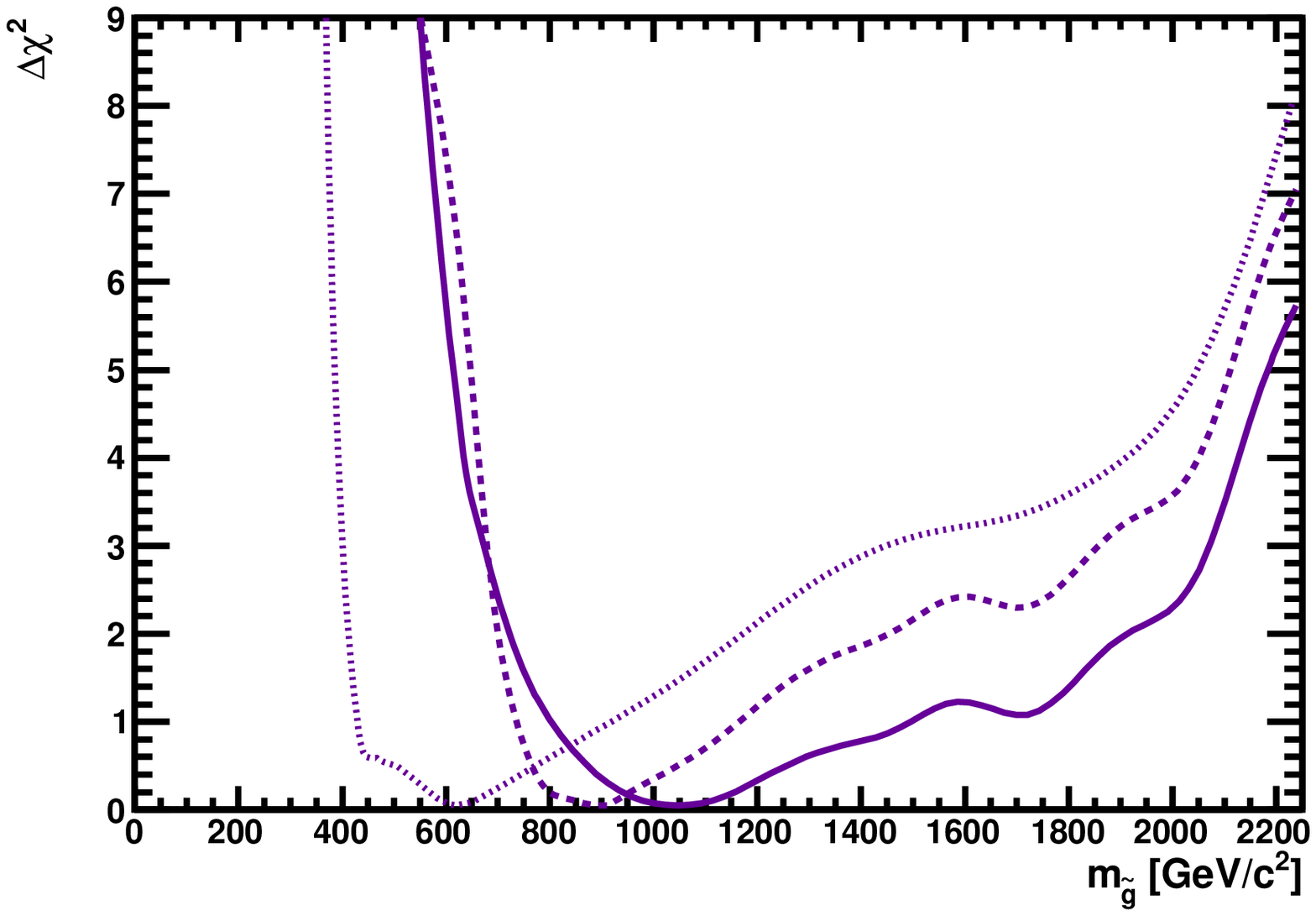}}
\resizebox{8cm}{!}{\includegraphics{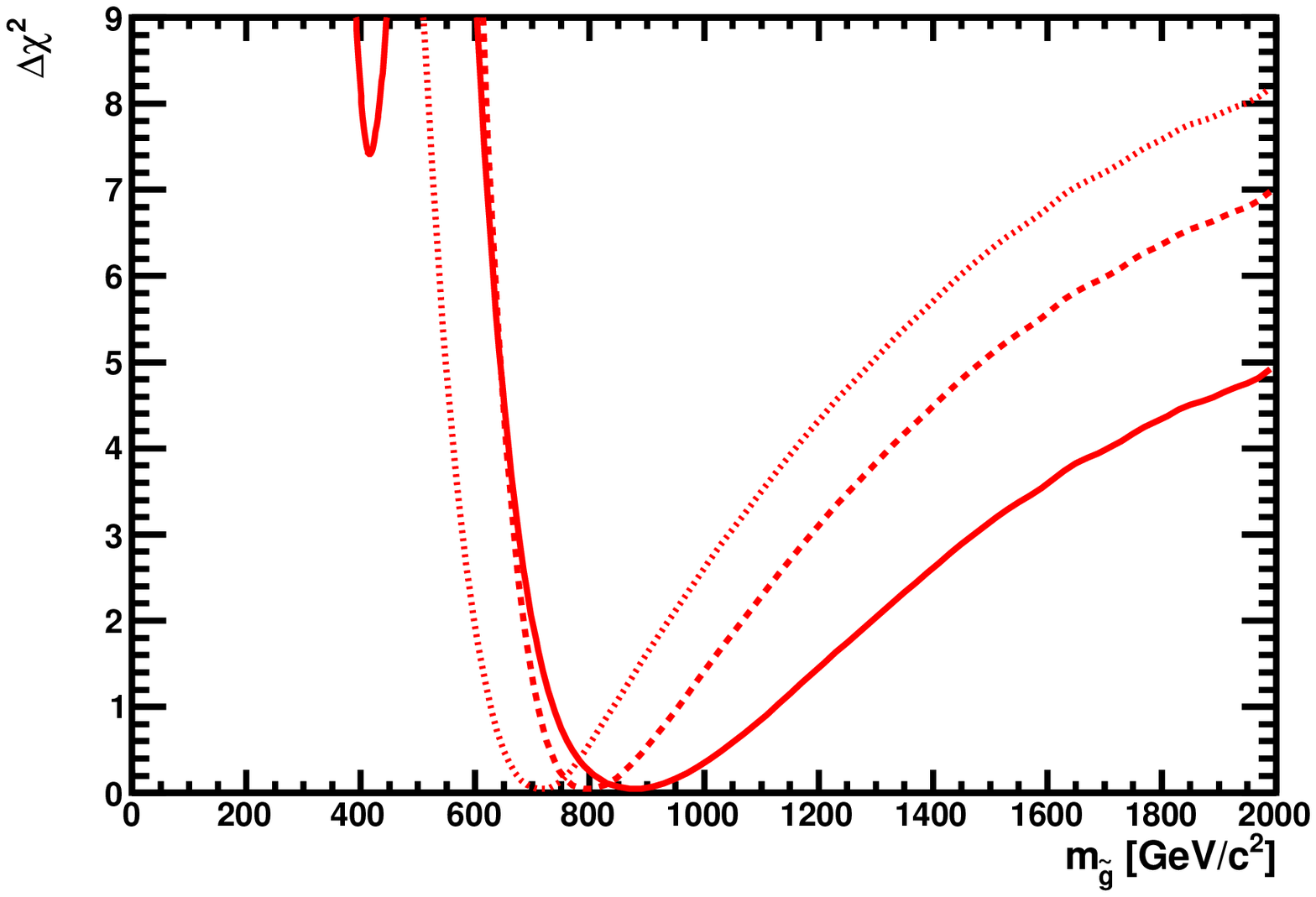}}
\resizebox{8cm}{!}{\includegraphics{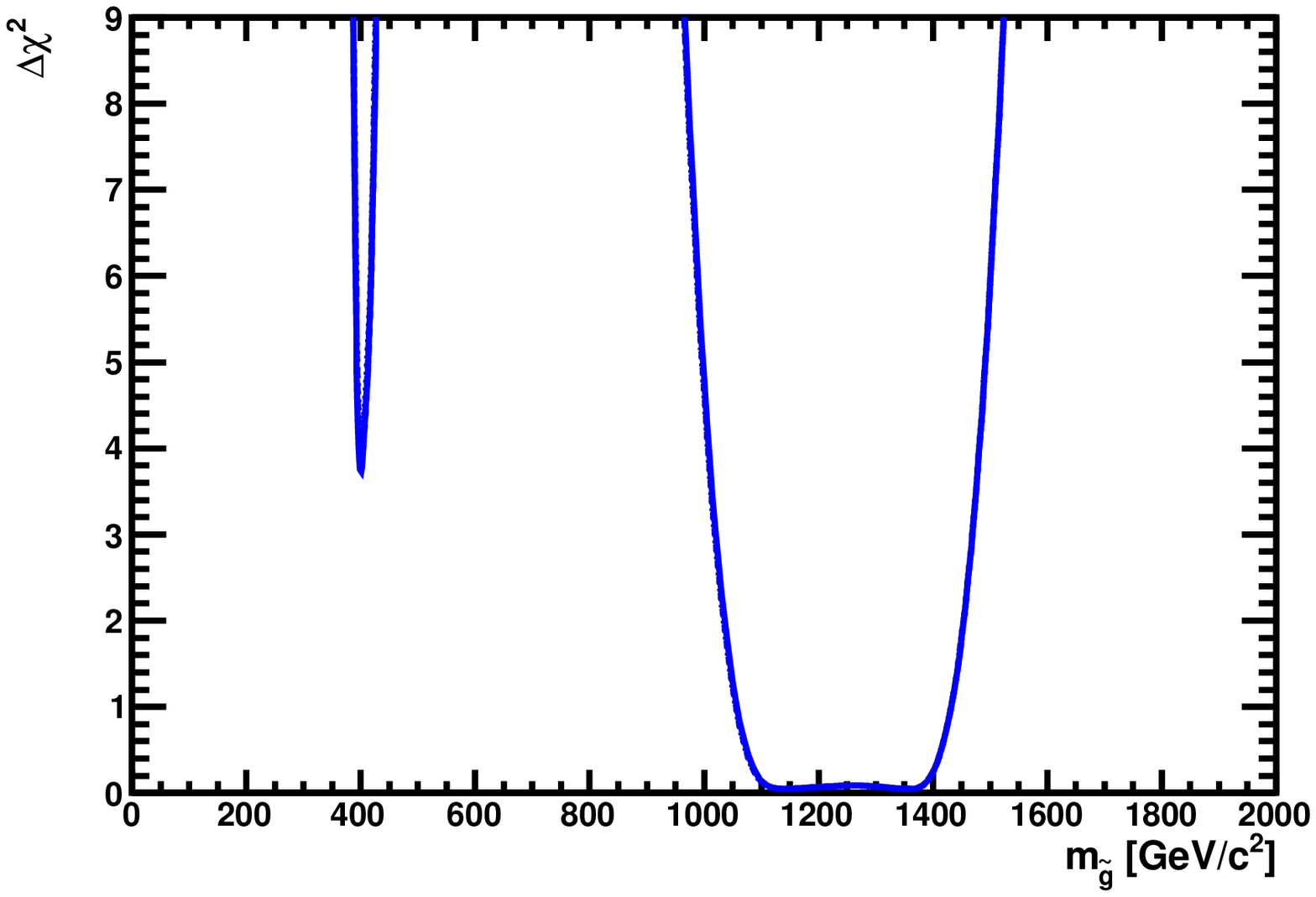}}\\
\caption{\it The one-parameter $\chi^2$ likelihood functions for the gluino
mass $\mgl$ in the CMSSM (upper left),
NUHM1 (top right), VCMSSM (lower left) and mSUGRA (lower right).
In each panel, we show  the $\chi^2$ function including the
CMS or ATLAS~\cite{CMSsusy,ATLASsusy} 
constraints as dashed and solid lines, respectively, and the previous
$\chi^2$ function as a dotted line.}
\label{fig:mgl}
\end{figure*}

In the region of parameter space of interest to the CMSSM, NUHM1, VCMSSM
and mSUGRA, the reach of the LHC for SUSY is largely determined by the
gluino mass, $\mgl$. Accordingly, we display in Fig.~\ref{fig:mgl} the
one-parameter $\chi^2$ functions for $\mgl$ relative to the minima in
all these models. In each  
case, we display the new likelihood functions incorporating CMS or ATLAS 
data as dashed or solid lines, respectively, and those given by the
previous fits as dotted lines 
(see also \reffis{fig:spectrum}, \ref{fig:spectrum2}), setting to zero 
the value of $\chi^2$ at the best fit in each model.
In this figure the one-parameter 
$\chi^2$ functions for mSUGRA are unchanged when the LHC data are
included but are shown for comparison purposes.

For each of the CMSSM, NUHM1 and VCMSSM, we see that the side of
the likelihood function below the best-fit point is shifted to larger $\mgl$
by similar amounts $\delta \mgl \sim 100$ to $200 \gev$. 
The best-fit values of $\mgl$ in the CMSSM, NUHM1 and VCMSSM are now
$\sim 900 \gev$.
In mSUGRA the most likely values of $\mgl$ are unchanged, lying in the range
$\sim 1000$ to $\sim 1400 \gev$, rising steeply outside this interval.


\section{Prediction for \boldmath{$\Mh$} including LHC data}
\label{sec:MhLHC}

\begin{figure*}[htb!]
\resizebox{8cm}{!}{\includegraphics{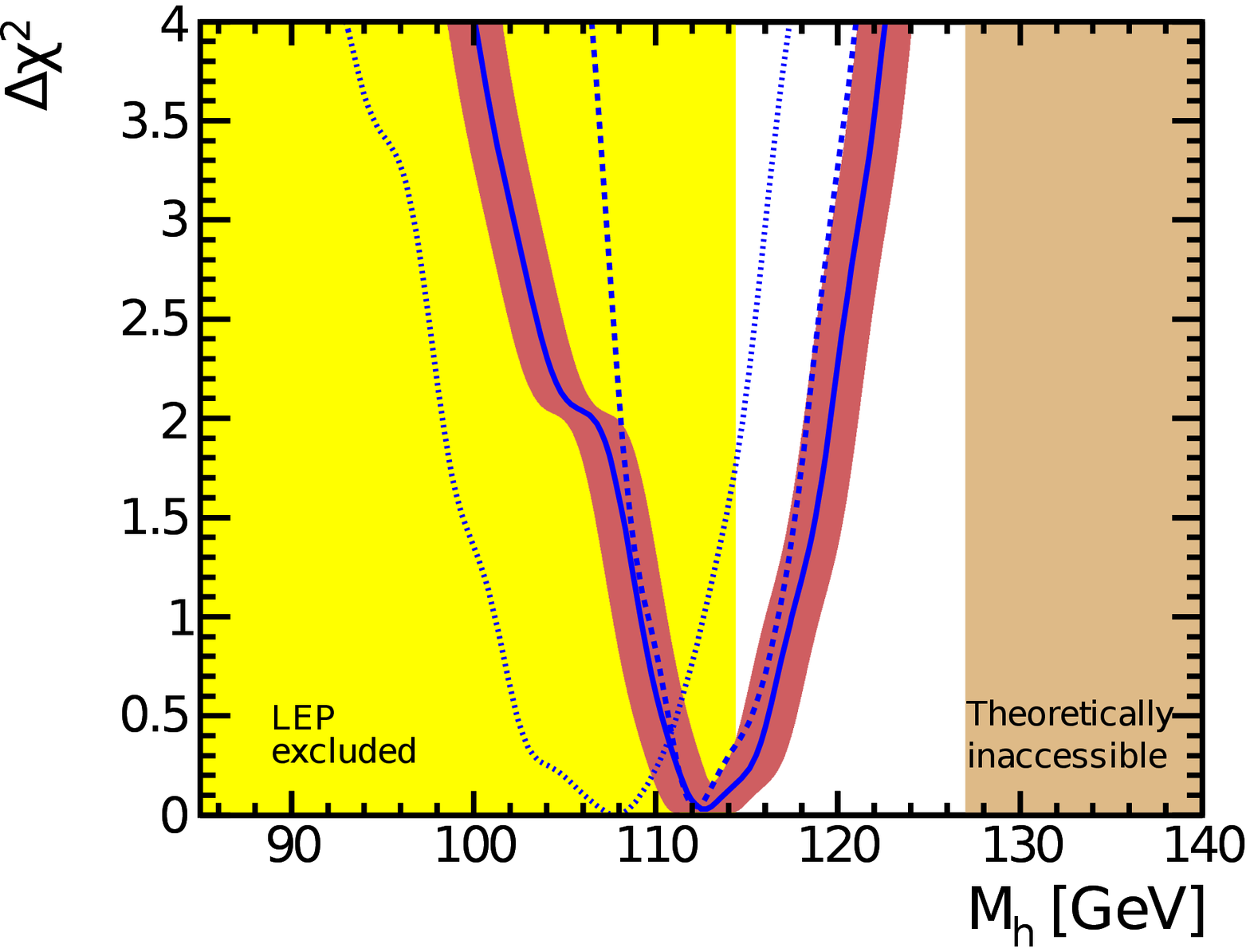}}
\resizebox{8cm}{!}{\includegraphics{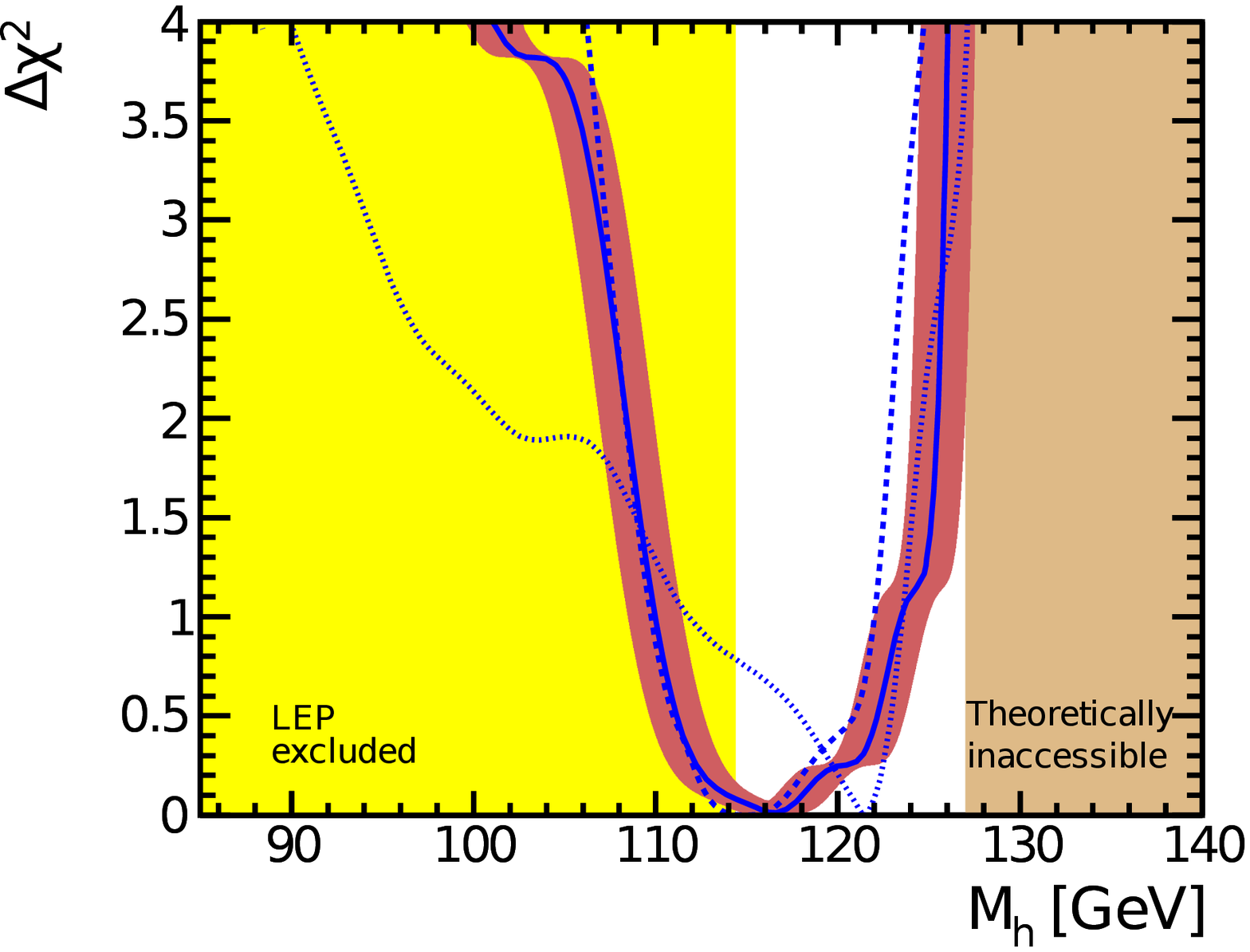}}
\resizebox{8cm}{!}{\includegraphics{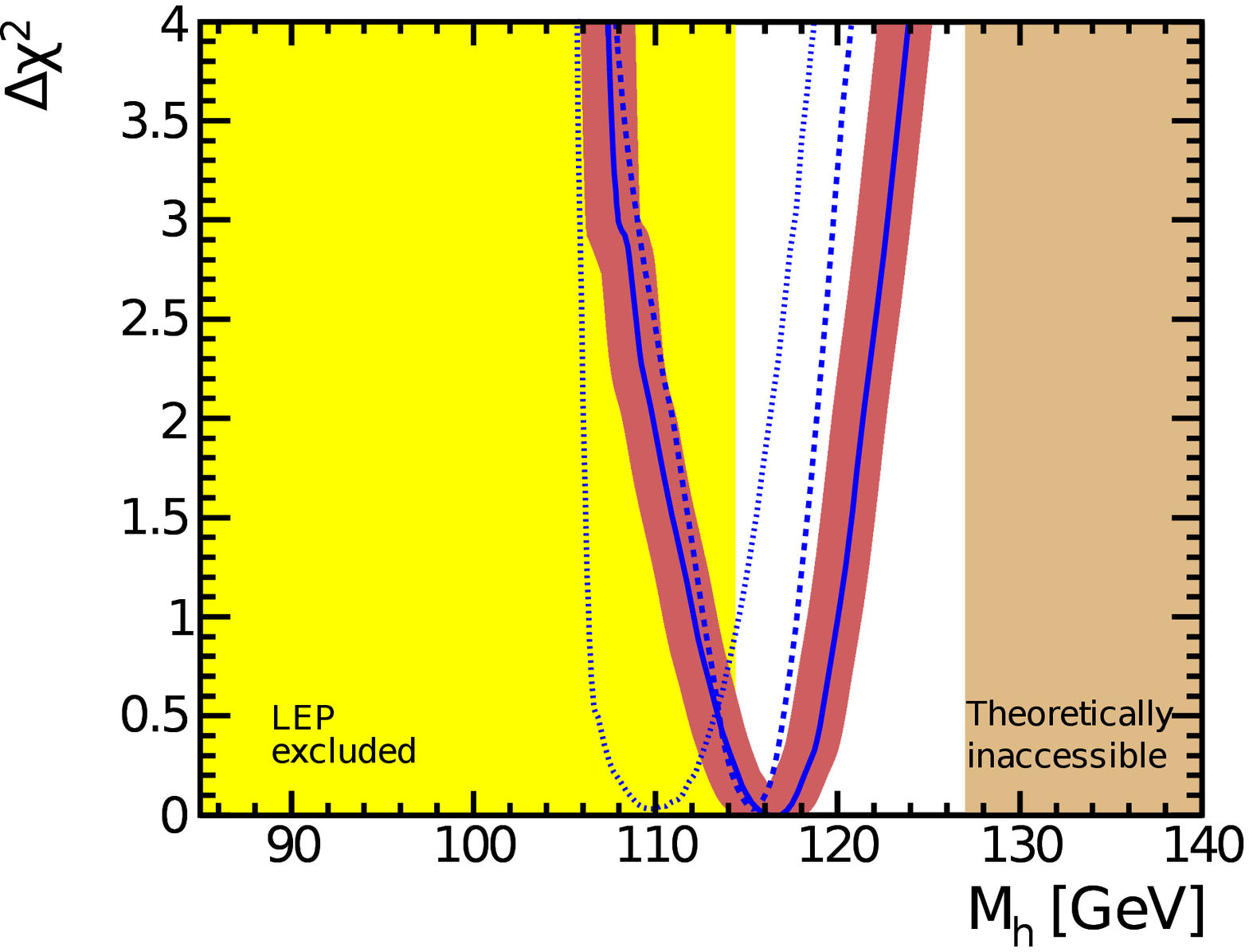}}
\resizebox{8cm}{!}{\includegraphics{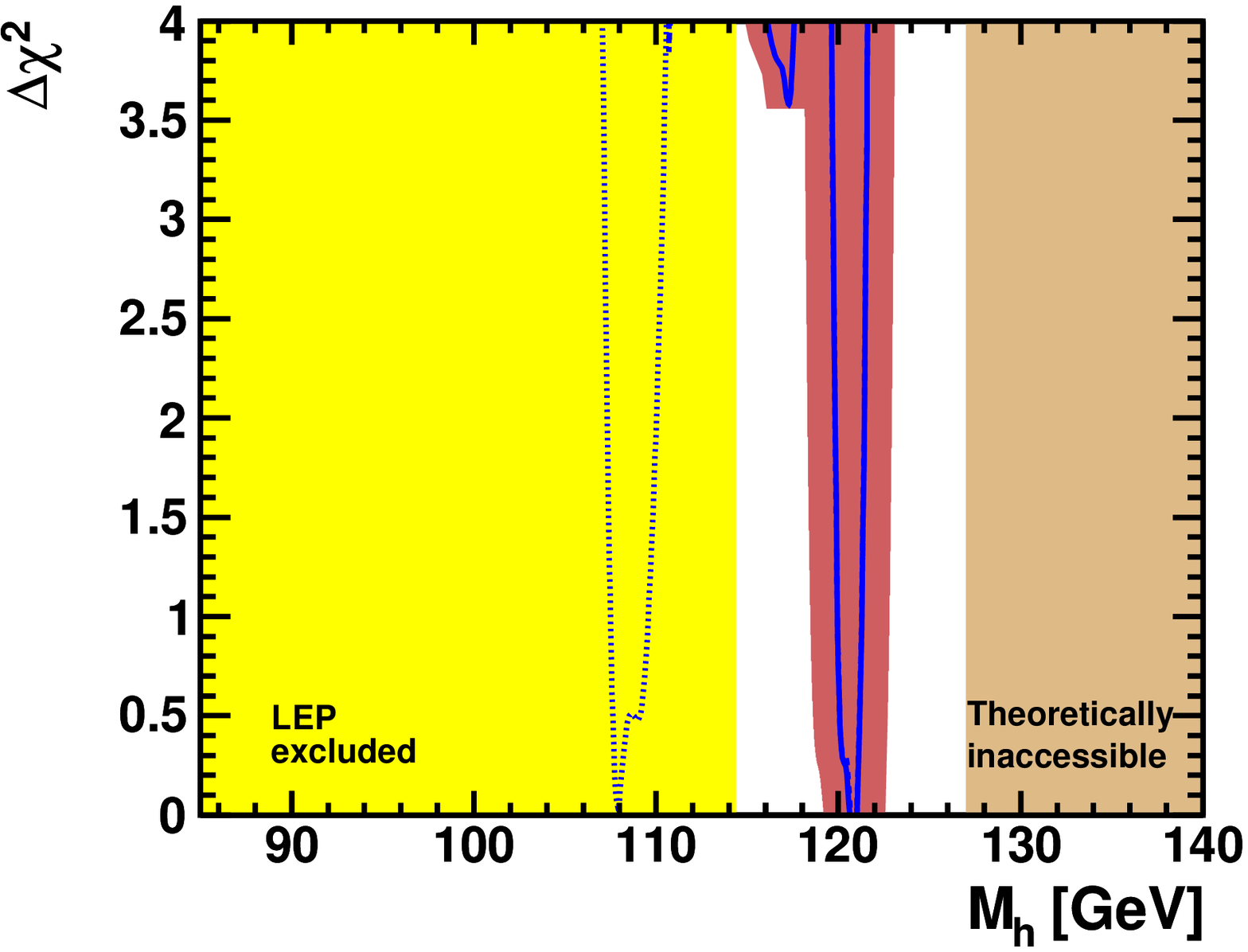}}\\
\vspace{-1cm}
\caption{\it The one-parameter $\chi^2$ likelihood functions for the
lightest MSSM Higgs mass $\Mh$ in the CMSSM (upper left),
NUHM1 (top right), VCMSSM (lower left) and mSUGRA (lower right). In each
panel, we show the $\chi^2$ functions including
the CMS (ATLAS)~\cite{CMSsusy,ATLASsusy} 
constraints as dashed (solid) lines, the latter with a red band
indicating the estimated theoretical uncertainty 
in the calculation of $\Mh$ of $\sim 1.5 \gev$, and the pre-LHC $\chi^2$
function is shown as a dotted line. 
}
\label{fig:mh}
\end{figure*}

One of the main goals of the LHC is the discovery of a Higgs boson,
revealing the mechanism of electroweak symmetry breaking. Similarly, a
huge effort is put in the current Tevatron analyses for SM and MSSM
Higgs boson searches.
Accordingly, we display in
Fig.~\ref{fig:mh} the one-parameter $\chi^2$ functions for 
the lightest MSSM Higgs mass $\Mh$ in the CMSSM, NUHM1, VCMSSM 
and mSUGRA. In this figure we do not include the direct limits from
LEP~\cite{Barate:2003sz,Schael:2006cr} or the Tevatron, so as to display
the (lack of) conflict between these 
limits and the predictions of supersymmetric models.
For each model we again display the new likelihood functions incorporating
CMS (ATLAS) data as dashed (solid) lines, around the latter indicating the
theoretical uncertainty in the calculation of $\Mh$ of $\sim 1.5 \gev$
by red bands. We also show, as dotted lines without red bands, the pre-LHC
predictions for $\Mh$ (again discarding the LEP constraint.)

In the case of the CMSSM, we see that the LHC constraints increases the
consistency of the model prediction with the direct LEP limit on $\Mh$,
indicated by the yellow region: the best-fit value can be found at 
$\sim 112.6 \gev$ 
with an estimated theoretical error of $1.5 \gev$, see also
\refta{tab:compare}.
In the case of the NUHM1, we see that the main effect of the LHC data is
to increase substantially the one-parameter $\chi^2$ function at low masses  
$\Mh < 110 \gev$, see \refse{sec:approach} for details.
In the NUHM1 the LEP constraint is weakened at low $\Mh$ because the
$hZZ$ coupling may be reduced (which is not possible in the CMSSM,
VCMSSM and mSUGRA~\cite{Ellis:2001qv,Ambrosanio:2001xb}). 
The LHC data render a large reduction less likely. Now most of the preferred
$\Mh$ region in the NUHM1 is indeed above $\sim 114 \gev$.
In the case of the VCMSSM, the LHC data strongly disfavor a
Higgs boson below the LEP limit.
In the case of mSUGRA, low $\Mh \sim 110$~GeV would have been favored pre-LHC,
but the CMS or ATLAS data push the preferred range to larger $\Mh$ compatible
with the LEP constraint~\cite{Barate:2003sz,Schael:2006cr}. 
Thus, including the LHC data, within NUHM1, VCMSSM and mSUGRA
the combination of all other experimental
constraints {\em naturally} evades the LEP Higgs constraints, and no
tension between $\Mh$ and the experimental bounds exists. Within the
CMSSM a ``tension'' smaller than in the SM can be observed.


\subsection*{Acknowledgements}

We thank O.~Buchm\"uller, R.~Cavanaugh, D.~Colling, A.~De~Roeck,
M.~Dolan, J.~Ellis, H.~Fl\"acher, G.~Isidori, K.~Olive, S.~Rogerson, 
F.~Ronga and G.~Weiglein with whom
the results presented here have been obtained.
We are grateful to the support of the
Spanish MICINN's Consolider-Ingenio 2010 Program under grant MultiDark
CSD2009-00064.
Finally, we thank the organizers of {\em  Kruger 2010: Workshop on
Discovery Physics at the LHC} for the invitation and an inspiring
workshop, as well as the elephant for showing up at the conference dinner.



\end{document}